\documentclass[twocolumn,twocolappendix]{aastex63}

\usepackage{color}
\usepackage{graphicx}
\usepackage{amsmath}    
\usepackage{amssymb}
\usepackage{mathrsfs}
\usepackage{times}
\usepackage{xcolor}
\usepackage{xspace}
\usepackage{datetime}
\usepackage{version}

\usepackage{natbib}

\newcommand{\RN}[1]{%
  \textup{\uppercase\expandafter{\romannumeral#1}}%
}

\newcommand{\simgt}{\lower.5ex\hbox{$\; \buildrel > \over \sim \;$}}
\newcommand{\simlt}{\lower.5ex\hbox{$\; \buildrel < \over \sim \;$}}

\newcommand{\Mpch}{\,h^{-1}\,\mathrm{Mpc}}
\newcommand{\Om}{\Omega_\mathrm{m}}
\newcommand{\OL}{\Omega_\Lambda}
\newcommand{\Ob}{\Omega_\mathrm{b}}
\newcommand{\gtot}{g_\mathrm{tot}}
\newcommand{\gbar}{g_\mathrm{bar}}
\newcommand{\rcut}{r_\mathrm{cut}}
\newcommand{\Mtot}{M_\mathrm{tot}}
\newcommand{\Mbar}{M_\mathrm{bar}}
\newcommand{\Mdm}{M_\mathrm{dm}}
\newcommand{\Ndata}{N_\mathrm{data}}
\newcommand{\LCDM}{$\Lambda$CDM\xspace}

\shorttitle{Testing the Collisionless Nature of Dark matter with the Cluster RAR}
\shortauthors{Tam et~al.}

\graphicspath{{./}}
\DeclareUnicodeCharacter{2212}{-}
\begin{document}

\title{Testing the Collisionless Nature of Dark Matter with the Radial Acceleration Relation in Galaxy Clusters}

\correspondingauthor{Sut-Ieng Tam}
\email{sitam@asiaa.sinica.edu.tw}

\author{Sut-Ieng Tam}
\affiliation{Academia Sinica Institute of Astronomy and Astrophysics (ASIAA), No.1, Sec. 4, Roosevelt Rd, Taipei
10617, Taiwan}
\author{Keiichi Umetsu}
\affiliation{Academia Sinica Institute of Astronomy and Astrophysics (ASIAA), No.1, Sec. 4, Roosevelt Rd, Taipei
10617, Taiwan}
\author{Andrew Robertson}
\affiliation{Jet Propulsion Laboratory, California Institute of Technology, 4800 Oak Grove Drive, Pasadena, CA 91109, USA}
\author{Ian G. McCarthy}
\affiliation{Astrophysics Research Institute, Liverpool John Moores University, 146 Brownlow Hill, Liverpool L53RF, UK}

\begin{abstract}
The radial acceleration relation (RAR) represents a tight empirical relation between the inferred total and baryonic centripetal accelerations, $\gtot=G\Mtot(<r)/r^2$ and $\gbar=G\Mbar(<r)/r^2$, observed in galaxies and galaxy clusters. The tight correlation between these two quantities can provide insight into the nature of dark matter. Here we use BAHAMAS, a state-of-the-art suite of cosmological hydrodynamical simulations, to characterize the RAR in cluster-scale halos for both cold and collisionless dark matter (CDM) and self-interacting dark matter (SIDM) models. SIDM halos generally have reduced central dark matter densities, which reduces the total acceleration in the central region when compared with CDM. We compare the RARs in galaxy clusters simulated with different dark matter models to the RAR inferred from CLASH observations. Our comparison shows that the cluster-scale RAR in the CDM model provides an excellent match to the CLASH RAR obtained by Tian et~al. including the high-acceleration regime probed by the brightest cluster galaxies (BCGs). By contrast, models with a larger SIDM cross-section yield increasingly poorer matches to the CLASH RAR. Excluding the BCG regions results in a weaker but still competitive constraint on the SIDM cross-section. Using the RAR data outside the central $r<100$~kpc region, an SIDM model with $\sigma/m=0.3$~cm$^{2}$~g$^{-1}$ is disfavored at the $3.8\sigma$ level with respect to the CDM model. This study demonstrates the power of the cluster-scale RAR for testing the collisionless nature of dark matter.
\end{abstract}


\section{Introduction} 
\label{sec:intro}

Based on modern cosmological studies \citep[e.g.][]{2013ApJS..208...19H}, dark matter is known to be the dominant matter component in the Universe, while its nature is still a mystery. In the current concordance cosmological model, $\Lambda$ Cold Dark Matter (CDM), large-scale structures formed hierarchically, with dark matter halos growing through a series of mergers of smaller halos as well as accretion. This standard model provides a good description of the observed large-scale structure. 

However, there are issues on smaller scales that are potentially challenging for the CDM model. For example, collisionless dark matter particles in CDM produce cuspy dark matter halos, where the density rises towards the halo center, which is inconsistent with the cored density profiles inferred for the halos hosting some dwarf galaxies. This is the so-called cusp--core problem \citep{1994ApJ...427L...1F, 1994Natur.370..629M}.

\cite{2000PhRvL..84.3760S} proposed a promising alternative to collisionless CDM, known as Self-Interacting Dark Matter (SIDM). SIDM was proposed to solve the small-scale problems with CDM, while preserving the successful predictions on large-scales in the \LCDM model. Elastic collisions of dark matter particles effectively smooth out the mass distribution at the center of halos, leading to a deviation from the cuspy density profile of CDM halos. In Figure~\ref{fig:density_profile}, we compare the density profiles of cluster-scale halos for different matter components, shown separately for simulations with five different dark matter runs (for details, see Section~\ref{sec:data}). The figure shows that the central dark matter density decreases with increasing dark matter scattering cross-section. The scattering rate of dark matter particles, $\Gamma$, is proportional to the local dark matter density $\rho_\mathrm{dm}(r)$, the dark matter scattering cross-section $\sigma$, and the local velocity-dispersion $v(r)$ of dark matter particles \citep{2013MNRAS.430...81R},
\begin{equation}
    \Gamma(r)\simeq\rho_\mathrm{dm}(r)v(r)\sigma/m,
    \label{eq:scatter}
\end{equation}
where $m$ is the dark matter particle mass. Therefore, with the highest $\rho_\mathrm{dm}$ and $v$, massive galaxy clusters are crucial laboratories to search for dark matter self-interactions. 

Galaxy clusters are the most massive gravitationally-bound structures resulting from the hierarchical formation process. About $85\%$ of their mass content is invisible dark matter, with the remainder being baryons that are mostly in the form of X-ray emitting hot gas. Since cluster properties depend on the growth of structure, they contain an abundance of cosmological and astrophysical information. Several characteristic features of galaxy clusters have been used to test dark matter models, such as dark matter halo shapes \cite[e.g.][]{2013MNRAS.430..105P,2018ApJ...860..104U}, offsets between dark matter and galaxies in merging systems \cite[e.g.][]{2015Sci...347.1462H,2018ApJ...869..104W, 2018MNRAS.477..669M}, the wobbling of brightest central galaxies \citep{2019MNRAS.488.1572H} and the amount and lensing efficiency of dark matter substructures \citep[e.g.][]{2016MNRAS.463.3876J,2020MNRAS.496.4032T, 2020Sci...369.1347M}. 

In the \LCDM model, dark matter has negligible interaction with baryons, except for gravity. However, tight relations between the distribution of dark matter and of baryonic matter have been discovered.
At the scale of spiral galaxies, the ratio of dynamical to baryonic masses, $\Mtot(<r)/\Mbar(<r)$, is found to be tightly coupled with gravitational acceleration, whereas no clear correlation with other physical quantities, such as galaxy size, has been found to date \citep{2004ApJ...609..652M}. 

By analyzing rotation curves of 153 spiral galaxies, \cite{2016PhRvL.117t1101M} found a tight correlation between two independent observables, namely the centripetal acceleration $\gtot(r) = V^2/r = G\Mtot(<r)/r^2$ and the baryonic contribution to this acceleration $\gbar(r)=GM_\mathrm{bar}(<r)/r^2$:
\begin{equation}
\frac{\gtot}{\gbar}=\frac{\Mtot}{\Mbar}=\frac{1}{1-e^{-\sqrt{\gbar/g_\dagger}}},
\label{eq:McGaugh}
\end{equation}
characterized by a characteristic acceleration scale, $g_{\dagger}= 1.20\pm0.24\times10^{−10}$~m~s$^{-2}$. This empirical relation between the total and baryonic centripetal accelerations is referred to as the radial acceleration relation (RAR). Since then, much effort have been invested to study the RAR in various galaxy samples \citep[e.g.,][]{2017ApJ...836..152L,2018MNRAS.477..230R,2019ApJ...877...18C, 2021A&A...650A.113B,2020arXiv200606700O}. Hydrodynamical simulations in the \LCDM framework have succeeded in reproducing the observed RAR of galaxies \citep[e.g.,][]{2017ApJ...835L..17K,2017PhRvL.118p1103L,2018PhRvL.120z1301G,2019MNRAS.485.1886D,2021MNRAS.507..632P}.

Recently, observational studies of the RAR have been extended to cluster-scale objects \citep[e.g.,][]{Tian2020,2020MNRAS.492.5865C,2021PDU....3100765P,2021PDU....3300854P,2022arXiv220501110E}. \cite{Tian2020} studied the RAR for a subsample of 20 high-mass galaxy clusters targeted by the CLASH program \citep{Postman2012}. In their analysis, the total mass of each cluster is inferred from a combined analysis of strong and weak lensing data \citep{Umetsu2016} and the baryonic mass from X-ray gas mass and stellar mass estimates \citep{Donahue2014}. \cite{2020MNRAS.492.5865C} analyzed X-ray data for a sample of 52 non-cool-core clusters. They obtained the cluster RAR using X-ray hydrostatic estimates for the total mass and X-ray gas mass estimates for the baryonic mass, ignoring the stellar mass contribution to the baryonic component. \cite{2022arXiv220501110E} studied the total and baryonic mass distributions for a sample of 12 X-COP clusters with X-ray and Sunyaev--Zel'dovich (SZ) effect observations, accounting for the stellar mass contribution. They found a complex shape of the RAR that strongly departs from the RAR in galaxies.

All of these studies found that the characteristic acceleration scale $g_{\dagger}$ in clusters is about an order of magnitude larger than that obtained from galaxy-scale objects. Observational RAR studies have also been extended to group-scale objects. \cite{2021PDU....3300874G} found that $g_{\dagger}$ of group-scale halos falls in between that found for galaxies and galaxy clusters. These observations suggest that there is no universal RAR that holds at all scales from galaxies to galaxy clusters.


Alternatively, the RAR observed at galaxy scales has been attributed to modified Newtonian dynamics \citep[MOND;][]{MOND}, which introduces a characteristic acceleration scale, $g_{\dagger}$, and modifies the dynamical law. However, MOND falls short in accounting for the total observed gravitational mass in galaxy clusters. For completeness it is worth mentioning other approaches to generalize the RAR to galaxy clusters, especially the eMOND framework of \cite{2012PhRvD..86f7301Z}, where the parameter $g_\dagger$ is allowed to be a monotonic increasing function of the system's escape velocity such that $g_\dagger$ is several times greater in clusters than in field galaxies. \cite{Zhao2017a} followed up this idea and presented fits to nearby relaxed clusters in the Chandra Sample \citep{2006ApJ...640..691V}. They succeed in this by tailoring $g_\dagger$ as a specific function of the Newtonian potential of the gas and the BCG galaxies (see their Eq. 20 and their Figs A.1-A.11 for 20 galaxy clusters). 


In this study, we use cosmological hydrodynamical simulations to study the RAR for simulated halos in both CDM and SIDM scenarios. We aim to explore a new method for constraining the collisionless nature of dark matter using the cluster-scale RAR, as well as to compare the RARs derived from numerical simulations with multiwavelength cluster observations.

This paper is organized as follows. Section~\ref{sec:data} introduces the simulation data sets we use in this work. Section~\ref{sec:result} shows the results of the halo RAR obtained from the simulations. Section~\ref{sec:observed} compares the theoretical predictions from the simulations with observational data from the CLASH program. In Section~\ref{sec:discussion}, we discuss the results and implications of our findings. Finally a summary is given in Section~\ref{sec:summary}.

Throughout this paper, we assume a Wilkinson Microwave Anisotropy Probe (WMAP) 9-year \LCDM cosmology \citep{2013ApJS..208...19H} with $\Om=0.287$, $\OL = 0.713$, and a Hubble constant of $H_0 = 100\,h$~km~s$^{-1}$~Mpc$^{-1}$ with $h=0.693$. We denote the critical density of the universe at a particular redshift $z$ as $\rho_\mathrm{c}(z)=3H^2(z)/(8\pi G)$ with $H(z)$ the Hubble function. We also define the dimensionless expansion function as $E(z)=H(z)/H_0$. We adopt the standard notation $M_\Delta$ to denote the total mass enclosed within a sphere of radius $r_\Delta$ within which the mean overdensity is $\Delta\times \rho_\mathrm{c}(z)$. We use ``$\ln$'' to denote the natural logarithm.

\section{Numerical Simulations} 
\label{sec:data}

\subsection{SIDM Models}
\label{subsec:sidm}

In this work, we use simulations run with four different SIDM models, as well as CDM, that were presented in \citet{2019MNRAS.488.3646R}. Three of the SIDM models have velocity-independent cross-sections with isotropic scattering of $\sigma/m = 0.1, 0.3$, and $1$~cm$^{2}$~g$^{-1}$, which we refer to as SIDM0.1, SIDM0.3, and SIDM1.0, respectively. The other SIDM model (hereafter vdSIDM) has a velocity-dependent and anisotropic cross-section. 

The vdSIDM differential cross-section is \citep{2017MNRAS.467.4719R,Robertson2021} 
\begin{equation}
    \frac{d\sigma}{d\Omega}=\frac{\sigma_0}{4\pi \left[1+(v^2/w^2)\sin^2\frac{\theta}{2}\right]^2}, 
     \label{eq:vdSIDM2}
\end{equation}
where $w$ is a characteristic velocity below which the scattering is approximately isotropic with $\sigma=\sigma_0$. For collision velocities greater than $w$, scattering becomes anisotropic (favoring scattering by small angles) and the cross-section decreases. Our vdSIDM model has  $\sigma_0/m = 3.04$~cm$^{2}$~g$^{-1}$ and  $w=560$~km~s$^{-1}$, which was chosen to reproduce the best-fit cross-section in \cite{2016PhRvL.116d1302K}. 

To understand the macroscopic behavior of anisotropic particle interactions, we can introduce the concept of momentum-transfer cross-section for vdSIDM \citep[e.g.,][]{2017MNRAS.467.4719R,Robertson2021}:
\footnote{In \citet{2017MNRAS.467.4719R}, $\sigma_T$ is referred to as the modified momentum-transfer cross-section and denoted as $\sigma_{\tilde{T}}.$}
\begin{equation}
    \sigma_T=2\int\!(1-|\cos\theta|)\frac{d\sigma}{d\Omega}d\Omega.
     \label{eq:vdSIDM2}
\end{equation}
During collisions, the amount of momentum transferred along the collision direction is given as $\Delta p=p(1-\cos\theta)$ with a scattering angle of $\theta$. When the velocity increases, the vdSIDM cross-section becomes more anisotropic, favoring scatter by small angles. Therefore, to better describe the effects of an anisotropic cross-section, $\sigma_T$ gives more weight to larger angle scattering which contributes larger amount of momentum transfer, while it downweights the small-angle scatter.  After integrating over the solid angle, we obtain the following expressions for the total cross-section ($\sigma_\mathrm{tot}$) and the momentum-transfer cross-section ($\sigma_T$):
\begin{equation}
  \begin{gathered}
    \sigma_\mathrm{tot}(v)=\frac{\sigma_0}{1+v^2/w^2},\\
    \sigma_T(v)=\sigma_0\frac{4w^4}{v^4}\left[2\ln\left(1+\frac{v^2}{2w^2}\right)-\ln\left(1+\frac{v^2}{w^2}\right)\right].
     \label{eq:vdSIDM_final}
  \end{gathered}
\end{equation} 
For a massive cluster with $M_{200}=10^{15}M_\odot$ at $z=0$ and for a typical relative velocity between particle pairs of $\langle v_\mathrm{rel}\rangle \sim \sqrt{GM_{200}/r_{200}}$, we obtain
$\sigma_\mathrm{tot}(\langle v_\mathrm{rel}\rangle) = 0.40$~cm$^2$~g$^{-1}$ and 
$\sigma_T(\langle v_\mathrm{rel}\rangle)=0.25$~cm$^2$~g$^{-1}$ for our vdSIDM model. We refer the reader to \cite{Robertson2021} for more details about the effective (velocity averaged) cross-section of vdSIDM halos.

\subsection{BAHAMAS Simulations}
\label{subsec:BAHAMAS}

We use $N$-body particle data from the BAryons And HAloes of MAssive Systems (BAHAMAS) suite of cosmological hydrodynamical simulations \citep[][]{2017MNRAS.465.2936M,2018MNRAS.476.2999M} with WMAP 9-year \citep{2013ApJS..208...19H} cosmology. BAHAMAS implements sub-grid models for star formation, and stellar and black hole feedback, and produces a good match to the observed stellar mass function, as well as the X-ray luminosities and gas mass fractions of galaxy groups/clusters. The simulations  occupy large periodic boxes, $400\Mpch$ on a side. For the SIDM simulations, we use the BAHAMAS-SIDM suite \citep[][]{2019MNRAS.488.3646R} which used the same initial conditions and sub-grid models as BAHAMAS, but included an implementation of dark matter scattering. The parameters associated with the galaxy formation physics used in BAHAMAS-SIDM, were kept the same as for the original BAHAMAS CDM simulation. The friends-of-friends algorithm \citep{1985ApJ...292..371D} with a linking length of $0.2$ times the mean inter-particle separation was run on each $z=0.375$ simulation output. From each simulation we extract the 10,000 most massive friends-of-friends groups, which have spherical-overdensity masses in the range of $12.5<\log_{10}[E(z)M_{200}/M_{\odot}]<15.3$.

\begin{figure*}[htbp]
 \centering
 \includegraphics[scale=0.8]{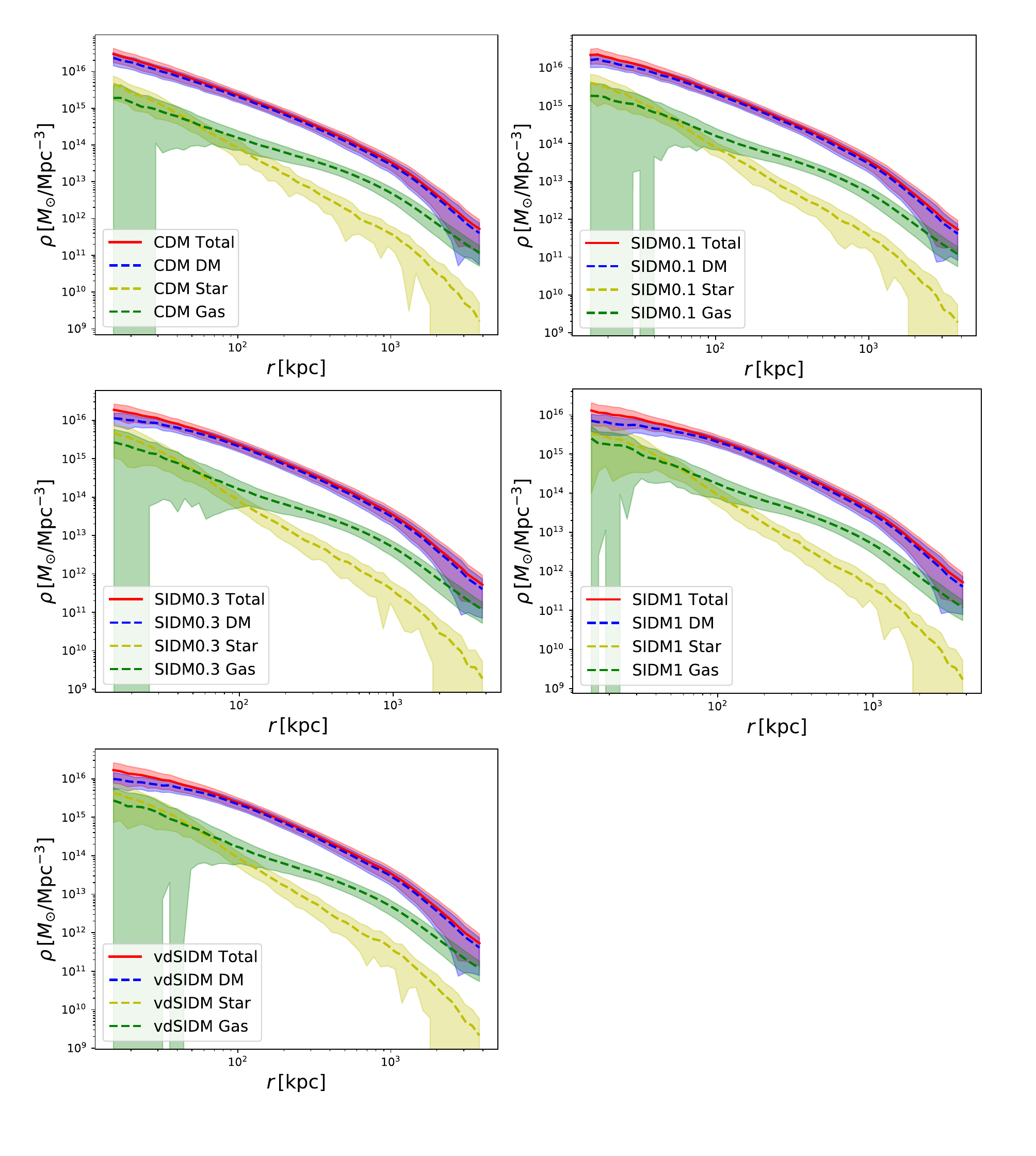}
 \caption{Mass density profiles for different matter components of cluster-scale halos with masses $E(z)M_{200} > 5\times 10^{14}M_\odot$ at $z=0.375$, shown for simulations with five different dark matter models. For each component, the dashed line represents the mean density profile and the shaded region shows the standard deviation around the mean profile. The large scatter in the innermost gas density is due to the AGN feedback in central galaxies, which removes most of the surrounding gas thus leading to low central gas densities for some halos.}
 \label{fig:density_profile}
\end{figure*}

For each halo we calculate the total enclosed mass profile, as well as the enclosed mass profile of the baryons. The center of the halo is defined by the location of the most gravitationally-bound particle, and the enclosed masses are calculated at 101 different radii, logarithmically spaced between proper (as opposed to comoving) lengths of $0.1$~kpc and $4$~Mpc. In addition, the total density as a function of radius is calculated by taking the difference in total enclosed mass at two successive radii, and dividing by the volume of the associated spherical shell. We consider the geometric mean of the inner and outer shell radii to be the radius at which this density is calculated. This density profile for each halo was used to make Figure~\ref{fig:density_profile}.

\section{Characterization of the Cluster-scale RAR in the BAHAMAS Simulations} 
\label{sec:result}

With the enclosed total and baryonic mass profiles $\Mtot(<r)=\Mdm(<r)+\Mbar(<r)$ and $\Mbar(<r)$ measured for each individual halo (Section~\ref{subsec:BAHAMAS}), we calculate their total and baryonic centripetal acceleration profiles as
\begin{equation}
\label{eq:g2M}
 \begin{aligned}
  \gtot(r) &= \frac{G\Mtot(<r)}{r^2},\\
  \gbar(r) &= \frac{G\Mbar(<r)}{r^2}.
 \end{aligned}
\end{equation}
Equation~(\ref{eq:g2M}) should be regarded as the definition of $\gtot(r)$ and $\gbar(r)$, not the result of assuming spherical symmetry. In this section, we aim to characterize the relationship between $\gtot$ and $\gbar$ for samples of halos selected from the BAHAMAS-CDM and -SIDM runs, focusing on massive cluster-scale objects.

\subsection{RARs in CDM and SIDM Halos}

\begin{figure*}[tbp]
 \centering
 \includegraphics[scale=0.9]{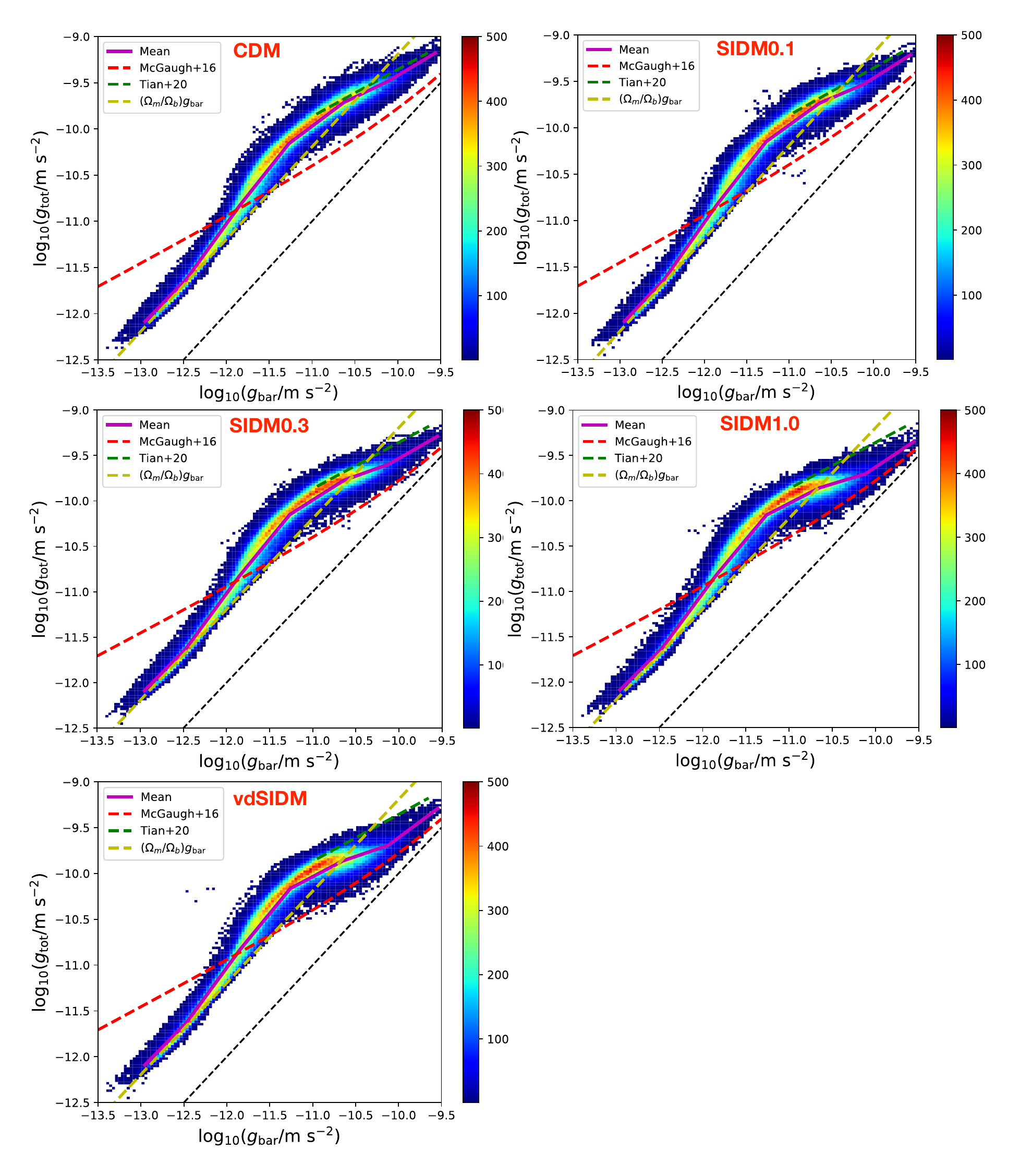}
 \caption{Histogram distribution of the baryonic and total centripetal accelerations ($\gbar, \gtot$) derived from the BAHAMAS halos at $z=0.375$ with masses $E(z)M_{200} > 5\times 10^{13} M_\odot$. Each panel shows the results for a different dark matter model. The halo accelerations are logarithmically extracted at scales from $r=15$~kpc to $r=4000$~kpc. In each panel, the magenta solid line represents the mean $\gtot$ as a function of $\gbar$ for that simulation. The red dashed line shows the \cite{2016PhRvL.117t1101M} relation observed in spiral galaxies (see Equation~(\ref{eq:McGaugh})). The yellow dashed line shows the linear relationship  $\gtot=(\Om/\Ob)\gbar$ corresponding to the cosmic ratio between total and baryonic mass densities. The green dashed line is the best-fit RAR observed for 20 high-mass galaxy clusters in the CLASH sample \citep{Tian2020}. The black dashed line shows the one-to-one relation.
 }
 \label{fig:all_RAR}
\end{figure*}

Figure~\ref{fig:all_RAR} shows the joint distribution of baryonic and total centripetal accelerations ($\gbar,\gtot$) derived from a subsample of group and cluster scale halos at $z=0.375$, with masses $E(z)M_{200}>5\times 10^{13}M_\odot$. The results are shown separately for the CDM and four different SIDM models. The halo centripetal accelerations are logarithmically sampled at scales from $r = 15$~kpc to $r = 4000$~kpc. 

For each dark matter run, the magenta solid line represents the mean $\gtot$ as a function of $g_\mathrm{bar}$ relation, or the halo RAR. The yellow dashed line represents the expectation corresponding to the cosmic mean ratio of total to baryonic mass densities, $\gtot=(\Om/\Ob)\gbar$. In all cases, the RARs of simulated halos converge towards the cosmic mean, $\gtot/\gbar=\Om/\Ob$, in the low-acceleration limit of $\gbar \simlt 10^{-13}$~m~s$^{-2}$. 

The red dashed line shows the \citet{2016PhRvL.117t1101M} relation (Equation~(\ref{eq:McGaugh})) observed in spiral galaxies over the acceleration range of $-12\simlt \log_{10}(\gbar/\mathrm{m~s}^{-2})\simlt -9$. Overall, the RARs of our simulated sample have a normalization that is higher than that observed at galaxy scales \citep{2016PhRvL.117t1101M}, suggesting a higher contribution from dark matter at a given baryonic acceleration.

\subsection{Mass Dependence of the Halo RAR}
\label{subsec:massdep}

\begin{figure}[tbp]
 \centering
 \includegraphics[scale=0.5]{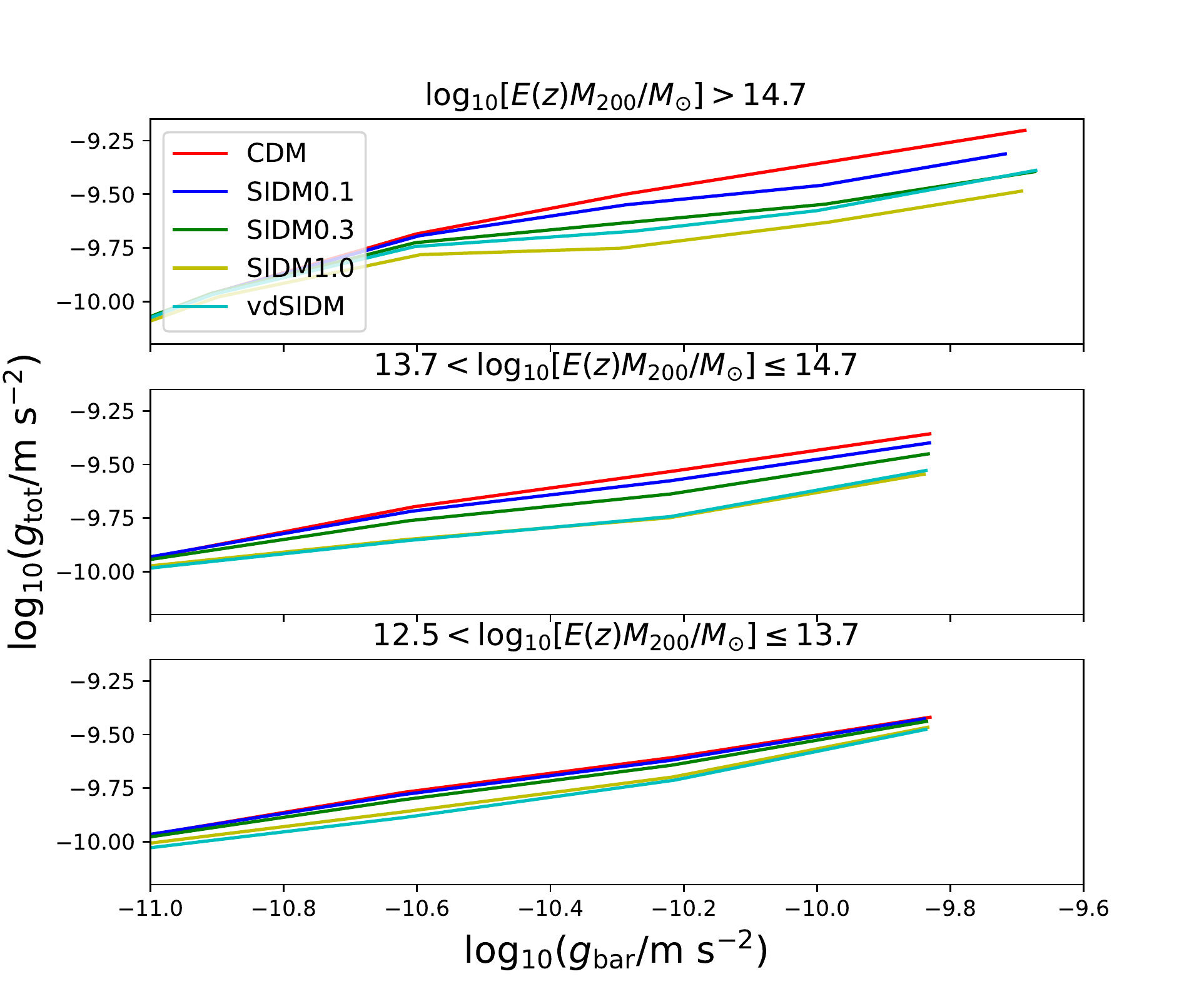}
 \caption{Comparison of the RARs for simulated halos in three different mass bins at $z=0.375$. In each panel, the results are shown for our five different dark matter models. A minimum cut-off radius of $\rcut=15$~kpc is used (see Figure~\ref{fig:all_RAR_radial_cut}).}
 \label{fig:all_RAR_mass_split}
\end{figure}

\begin{figure*}[tbp]
 \centering
 \includegraphics[scale=0.5]{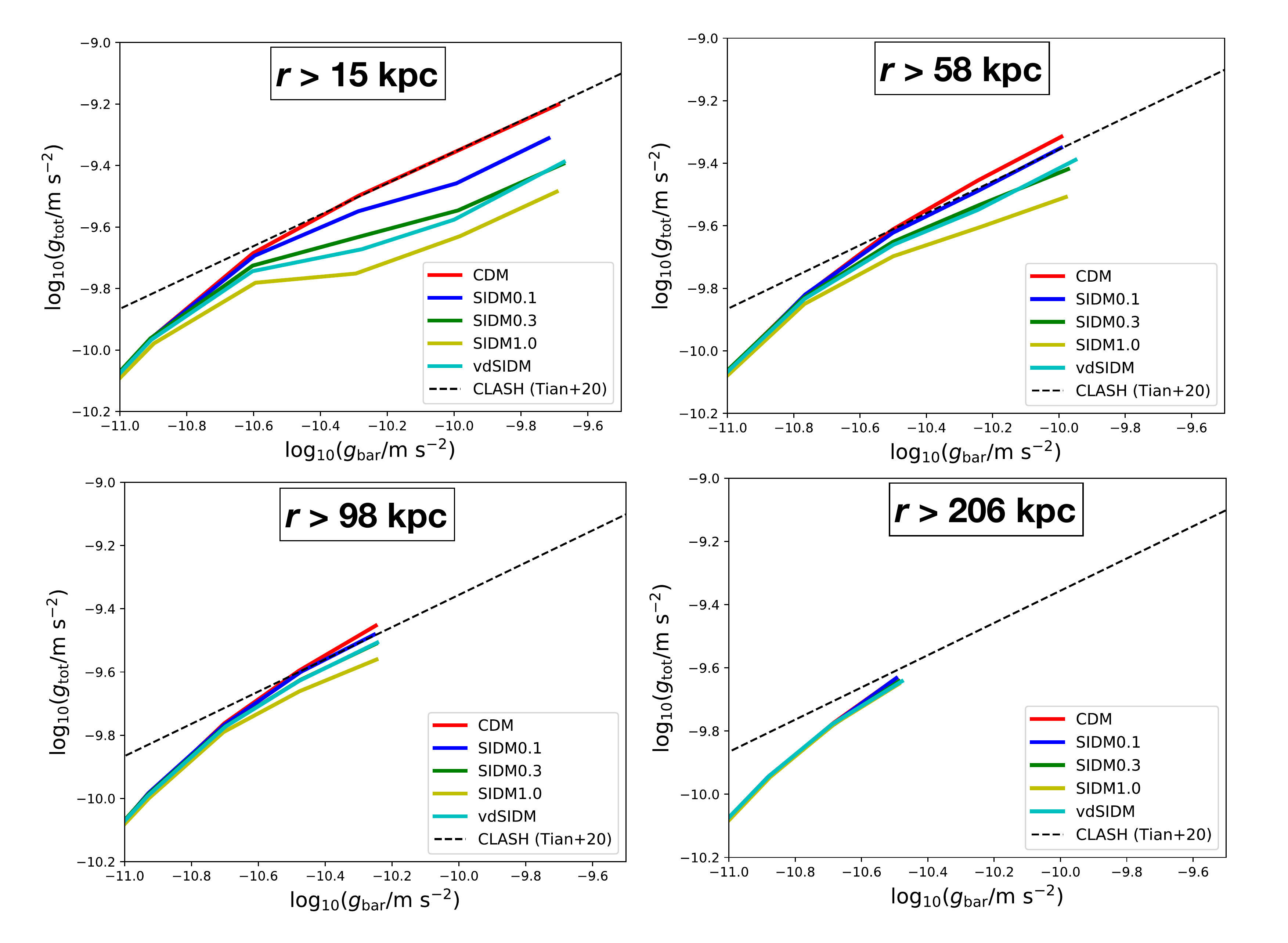}
 \caption{Dependence of the cluster-scale RAR on the inner radial cut. The four panels show the comparison of the RARs obtained with different radial cuts. Solid lines with different colors show the mean relations for different dark matter runs, derived for massive cluster halos with $E(z)M_{200}>5\times 10^{14}M_\odot$.}
 \label{fig:all_RAR_radial_cut}
\end{figure*}

The full sample of 10,000 BAHAMAS simulated halos spans a wide range of halo mass. We therefore split our sample into three mass bins: $12.5<\log_{10}[E(z)M_{200}/M_{\odot}] \le 13.7$, $13.7<\log_{10}[E(z)M_{200}/M_{\odot}]\le 14.7$, and $\log_{10}[E(z)M_{200}/M_{\odot}]>14.7$, to investigate the halo mass dependence of the RAR.

Figure~\ref{fig:all_RAR_mass_split} shows the resulting RARs in the three mass bins, for the five different dark matter models. For the lowest mass bin, the RARs for different dark matter models largely overlap with each other, while for the highest mass bin (corresponding to massive cluster halos) the slope of the RAR at high $\gbar$ decreases with increasing  SIDM cross-section. The flattening feature in $\gtot$ at high acceleration corresponds to the approximately constant-density dark matter ``cores'' at the center of SIDM halos. This distinguishing feature, being more significant in more massive halos, is consistent with the fact that the scattering rate is proportional to the local dark matter density and velocity dispersion (Equation.~(\ref{eq:scatter})). This result indicates that the high-acceleration cluster-scale RAR can be used to probe the nature of dark matter. In the following analyses, we will focus on the 48 cluster-scale halos in the highest mass bin, which are more sensitive to the SIDM cross-section. 

The velocity dependence of the vdSIDM cross-section is apparent in Figure~\ref{fig:all_RAR_mass_split}. For high mass halos, vdSIDM behaves most similarly to SIDM0.3 while for the group-scale halos, vdSIDM halos are most like SIDM1.0 halos. This behaviour is consistent with \citet[][see their Figure~1]{2019MNRAS.488.3646R}, and reflects the fact that the vdSIDM cross-section decreases with increasing relative velocity between dark matter particles, and relative velocities are larger in more massive halos.

In the above analyses, a minimum radius of $r=15$~kpc is applied. From cluster observations, the enclosed mass in the inner regions $r\in [15,100]$~kpc is difficult to measure. We therefore study the RARs for the same sets of halos discussed above, but with three larger minimum radii. 

In Figure~\ref{fig:all_RAR_radial_cut}, we plot the RARs for the high-mass objects with radial cuts of $\rcut=15$~kpc,  $58$~kpc,  $98$~kpc, and  $206$~kpc, respectively. The radial cut of $98$~kpc corresponds approximately to the regime of combined strong and weak-lensing analyses, while a radial cut of $206$~kpc represents the regime of weak-lensing-only analyses for clusters at $z\simgt 0.1$. We recall that beyond $r\sim100$~kpc, it is challenging to distinguish the CDM and SIDM models by measuring the mass density profile of galaxy clusters, as shown in Figure~\ref{fig:density_profile}. The bottom-left panel of Figure~\ref{fig:all_RAR_radial_cut} shows that even beyond $r\sim 100$~kpc, the slope of the RAR for SIDM1.0 is shallower than that for CDM. Deviations in the RAR between CDM and SIDM is thus more significant than the conventional cusp--core features in the density profiles at larger radii. For the radial range of weak-lensing-only measurements, the discrepancy becomes tiny and would be almost impossible to detect. Hence, a combined strong and weak-lensing analysis is typically required to distinguish SIDM and CDM in terms of the RAR, because this enables the measurement of the total enclosed mass down to and below approximately $100$~kpc. 

The key features used to distinguish different dark matter models are largely driven by information in the inner region of cluster halos. In Appendix~\ref{app_1}, we further investigate whether the correlation between the acceleration ratio $\gtot/\gbar$ and the clustercentric distance $r$, namely the $\gtot/\gbar$--$r$ relation, can provide sufficient information to distinguish between different dark matter models. We find that the radial $\gtot(r)/\gbar(r)$ (or $\Mtot(r)/\Mbar(r)$ profiles of cluster halos derived from different dark matter runs of the BAHAMAS simulation significantly overlap with each other, even in their innermost region, and hence compared to the $\gtot$--$\gbar$ relation, the $\gtot/\gbar$--$r$ relation has a much weaker sensitivity to the SIDM cross-section. 

\subsection{Power-law Characterization of the Cluster-scale RAR}
\label{subsec:scaling}

\begin{deluxetable}{lccc}[tbp]
\centering
\tablecaption{Cluster-scale RAR and its intrinsic scatter characterized in the high-acceleration regime}
\label{tab:slope}
\tablehead{
 \colhead{Dark matter model} & 
 \colhead{{$b_1$}}& 
 \colhead{{$b_0$}} &
 \colhead{{$\Delta_\mathrm{int}$}}\\
 \colhead{} &
 \colhead{} &
 \colhead{} &
 \colhead{(dex)}
 }
\startdata
CDM &  0.526 & -4.096 & 0.064 \\
SIDM0.1 &0.422 & -5.225 & 0.058\\
SIDM0.3 & 0.350 & -6.025 & 0.076\\
SIDM1.0  & 0.334 & -6.270 & 0.100\\
vdSIDM & 0.379 &  -5.753 & 0.091
\enddata
\tablecomments{For each model, the RAR is derived for a subsample of cluster halos at $z=0.375$ with masses $E(z)M_{200}>5\times 10^{14}M_\odot$.  A cut-off radius of $\rcut=15$~kcp is used. The quantities $b_1$ and $b_0$ represent the slope and intercept of the power-law fit (Equation~(\ref{eq:powerlaw})) in the high-acceleration region of $\log_{10}(g_{\mathrm{bar}}/\mathrm{m~s}^{-2})>-10.6$. The $\Delta_\mathrm{int}$ parameter is the intrinsic scatter in dex.}
\end{deluxetable}

We characterize the halo RARs obtained in Section~\ref{subsec:massdep}, focusing on cluster halos in the highest-mass bin with $E(z)M_{200} > 5\times 10^{14}M_\odot$ at $z=0.375$. To this end, we assume a power-law function of the form:
\begin{equation}
  \log_{10}(\gtot/\mathrm{m~s}^{-2})=b_1\log_{10}(\gbar/\mathrm{m~s}^{-2})+b_0,
  \label{eq:powerlaw}
\end{equation}
with $b_1$ the logarithmic slope and $b_0$ the intercept. At $\gbar \sim 10^{-11}$~m~s$^{-2}$, the mean cluster RARs for all the dark matter runs converge to a power-law with $b_1\approx 1.15$ and $b_0\approx 2.68$ (see the top panel of Figure~\ref{fig:all_RAR_mass_split}). Then, the logarithmic slope begins to flatten gradually at $\gbar\simgt 10^{-11}$~m~s$^{-2}$. At $\gbar\simgt 10^{-10.6}$~m~s$^{-2}$, the cluster RAR is found to be highly sensitive to the SIDM cross-section.

In Table~\ref{tab:slope}, we summarize the best-fit values of $b_1$ and $b_0$ for the CDM and four SIDM runs characterized in the high-acceleration regime $\gbar > 10^{-10.6}$~m~s$^{-2}$. It should be noted that we also fitted the mean RARs at $\gbar>10^{-10.6}$~m~s$^{-2}$ using the \citet{2016PhRvL.117t1101M} relation (Equation~(\ref{eq:McGaugh})), finding that only the CDM case can be well described by this functional form with an acceleration scale of $g_{\dagger} = (1.42\pm0.06)\times 10^{-9}$~m~s$^{-2}$, which is much higher than the characteristic acceleration scale $g_{\dagger}\approx 1.2\times 10^{-10}$~m~s$^{-2}$ observed at galaxy scales (Section~\ref{sec:intro}).


Table~\ref{tab:slope} also lists the levels of intrinsic scatter ${\Delta_\mathrm{int}}$ around the mean relations obtained for the five different dark matter runs. In all cases, we find a remarkably tight distribution in $\log\gtot$--$\log\gbar$ space, with a slight increase in ${\Delta_\mathrm{int}}$ with increasing cross-section.
For the CDM, SIDM0.1 and SIDM0.3 models, the values of ${\Delta_\mathrm{int}}$ agree within the errors with {$\Delta_\mathrm{int}$}$ = 0.064^{+0.013}_{-0.012}$ (in dex) determined for the CLASH sample \citep{Tian2020}.

\subsection{Evolution of the Cluster-scale RAR}

\begin{figure}[tbp]
 \centering
 \includegraphics[width=0.5\textwidth,clip]{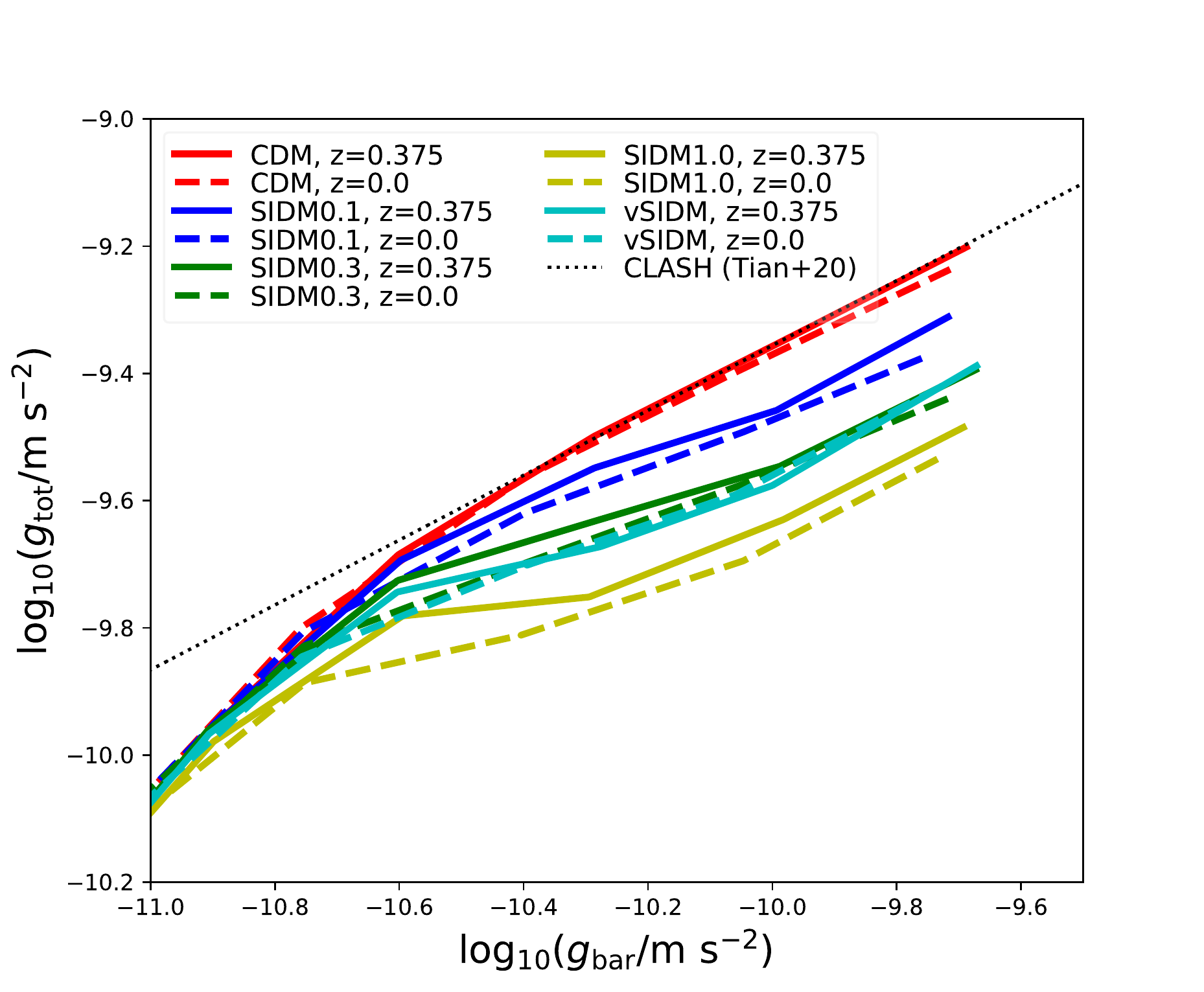}
 \caption{Comparison of the cluster-scale RARs between two different simulation snapshots, $z=0.375$ (solid line) and $z=0.0$ (dashed line), obtained for massive cluster halos with $E(z)M_{200}>5\times 10^{14}M_\odot$. The results are shown for five different dark matter runs. The dotted line shows the RAR derived for the CLASH sample at a median redshift of $\overline{z}=0.377$ \citep{Tian2020}.}
 \label{fig:compare_z}
\end{figure}

Here we investigate the evolution of the RAR by analyzing simulated halos at two redshifts, $z=0$ and $z=0.375$. For this purpose, we focus on massive cluster halos with $E(z)M_{200} > 5\times 10^{14}M_\odot$ (Sections~\ref{subsec:massdep} and \ref{subsec:scaling}). At $z=0$, we have a total of 82 cluster halos in this subsample. 

Figure~\ref{fig:compare_z} shows the comparison of the cluster-scale RARs at $z=0$ and $z=0.375$. Compared to $z=0.375$, we find a larger discrepancy at $z=0$ between the CDM and SIDM results: the larger the scattering cross-section, the lower the total acceleration at high $\gbar$ (see also Section~\ref{subsec:massdep}). 
The SIDM dependence increasing toward lower redshift is expected, because the radius out to which SIDM significantly affects density profiles is well described by the radius where there has been one scattering event per particle over the age of the halo \citep{Robertson2021}. This suggests that cluster RAR measurements for lower-$z$ samples should provide a more sensitive test of the SIDM cross-section. We find that the cluster RAR derived from vdSIDM at $z=0$ resembles well that from SIDM0.3 at $z=0$.

\section{Comparison with observational data} 
\label{sec:observed}

\begin{figure}[tbp]
 \centering
 \includegraphics[width=0.5\textwidth,clip]{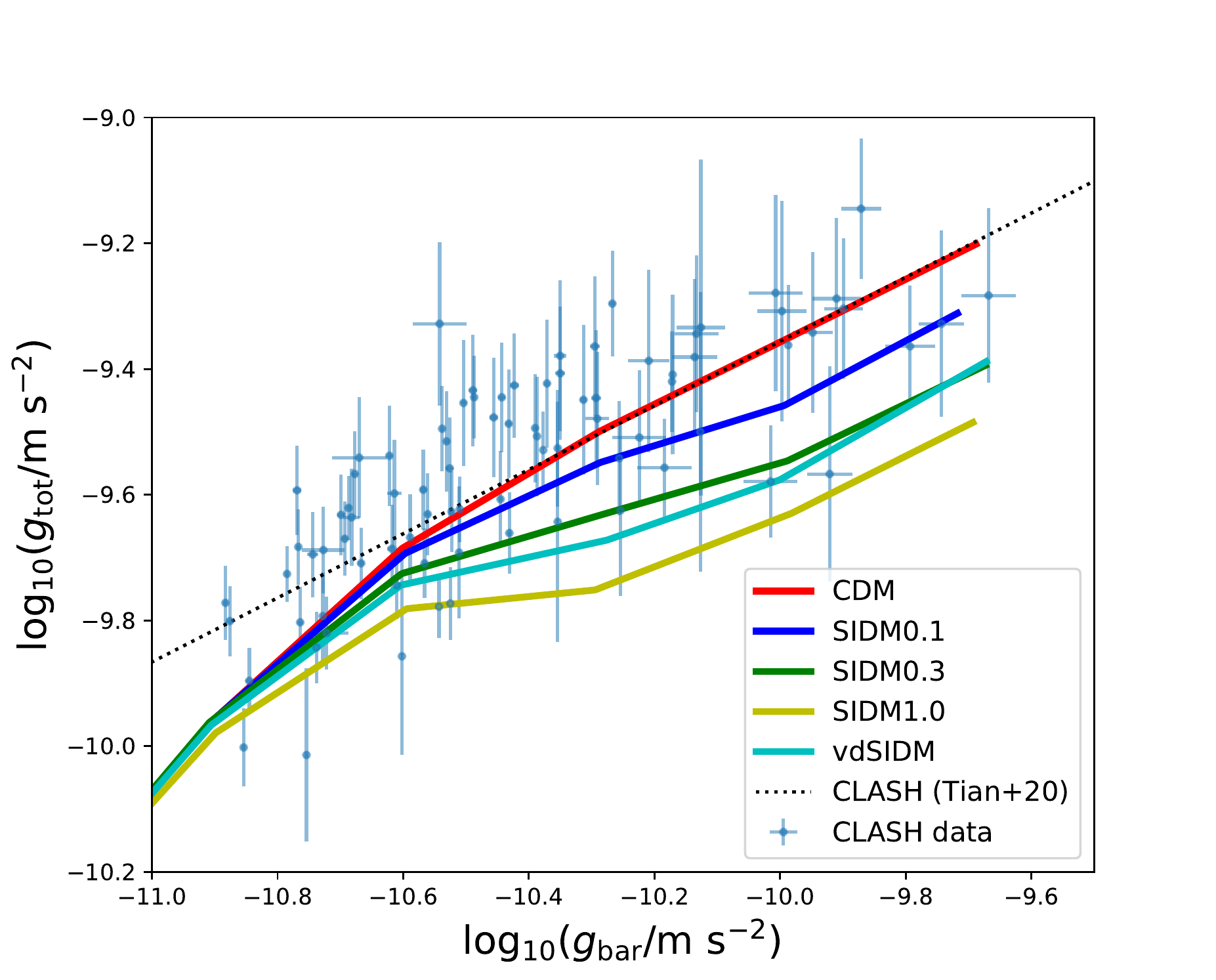}
 \caption{Comparison of the CLASH RAR \citep{Tian2020} with our BAHAMAS predictions of the cluster-scale RAR at $z=0.375$, derived from five different dark matter runs (thick color lines). The CLASH measurements of cluster centripetal accelerations ($g_\mathrm{bar}, g_\mathrm{tot}$) \citep{Tian2020} are shown with blue points with error bars. The black dotted line shows the best-fit RAR for the CLASH sample at a median redshift of $\overline{z}=0.377$.}
 \label{fig:CLASH}
\end{figure}

In this section, we compare the cluster-scale RARs derived from the BAHAMAS simulations with observations of galaxy clusters. An observational determination of the cluster RAR relies on accurate measurements of both total and baryonic mass profiles, $M_\mathrm{tot}(<r)$ and $M_\mathrm{bar}(<r)$, for a sizable sample of galaxy clusters. The baryonic mass in galaxy clusters is dominated by the hot intracluster gas, except in their central region where the BCG dominates the total baryonic mass \citep[e.g.,][]{Sartoris2020}.\footnote{The mean effective (half-light) radius of the CLASH BCGs is $\langle R_\mathrm{e}\rangle\sim 30$~kpc \citep{Tian2020}.}. Thus, baryonic mass estimates for both components are essential. Furthermore, unbiased estimates for the total mass in galaxy clusters are critical for a robust determination of the cluster RAR. 

In this context, \cite{Tian2020} determined the RAR at BCG--cluster scales for a sample of 20 high-mass galaxy clusters targeted by the CLASH program, by combining weak and strong gravitational lensing data \citep{Umetsu2014,Umetsu2016,Merten2015,Zitrin2015}, X-ray gas mass measurements \citep{Donahue2014}, and BCG stellar mass estimates. The stellar mass contribution from member galaxies was statistically corrected for. To date, this is the only study that uses gravitational lensing data to directly probe the total acceleration $g_\mathrm{tot}$ in galaxy clusters. By contrast, other studies based on hydrostatic estimates for $\gtot$ could potentially bias the true underlying RAR. In this study, we thus focus on the CLASH RAR studied by \citet{Tian2020}.
 
\subsection{CLASH Data}
\label{subsec:clash}

With the aim of precisely determining the mass profiles of galaxy clusters using deep 16-band imaging with the Hubble Space Telescope \citep[HST;][]{Postman2012} and ground-based weak-lensing observations \citep{Umetsu2014}, a subsample of 20 CLASH clusters was X-ray selected to be massive ($>5$~keV), with nearly concentric X-ray isophotes and a well-defined X-ray peak located close to the BCG. For this subsample, no lensing information was used a priori to avoid a biased sample selection. Cosmological hydrodynamical simulations suggest that the CLASH X-ray-selected subsample is mostly composed of relaxed systems ($\sim 70\%$) and largely free of orientation bias \citep{Meneghetti2014}. Another subsample of five clusters was selected by their exceptional lensing strength to magnify galaxies at high redshift. These clusters often turn out to be dynamically disturbed massive systems \citep{Umetsu2020}.

The CLASH sample spans nearly an order of magnitude in mass, $5\simlt M_{200}/10^{14}M_\odot\simlt 30$. For each of the 25 clusters, HST weak- and strong-lensing data products are available in their central regions \citep{Zitrin2015}.  \citet{Umetsu2016} combined wide-field weak-lensing data obtained primarily with Suprime-Cam on the Subaru telescope \citep{Umetsu2014} and the HST weak- and strong-lensing constraints of \citet{Zitrin2015}. For an observational determination of the RAR, \citet{Tian2020} combined the X-ray data products from \citet{Donahue2014} and the lensing data products from \citet{Umetsu2016}, yielding a subsample of 20 CLASH clusters composed of 16 X-ray-selected and 4 lensing-selected systems. We note that five clusters of the CLASH sample were not included in the joint weak- and strong-lensing analysis performed by \citet{Umetsu2016} because of the lack of usable wide-field weak-lensing data. Consequently, they were also excluded in our analysis. 

The CLASH subsample analyzed by \citet{Tian2020} has a median redshift of $\overline{z} = 0.377$, which closely matches our simulation snapshot at $z=0.375$. The typical resolution limit of the mass reconstruction set by the HST lensing data is $10\arcsec$, which corresponds to $\approx 50$~kpc at $\overline{z}=0.377$ \citep{Umetsu2016}. It was found by \citet{Umetsu2016} that the stacked lensing signal of the CLASH X-ray-selected subsample is well described by a family of cuspy, sharply steepening density profiles, such as the Navarro--Frenk--White \citep[][hereafter NFW]{NFW1996,NFW1997}, Einasto \citep{Einasto}, and DARKexp \citep{DARKexp} profiles. Of these, the NFW model best describes the CLASH lensing data \citep[see also][]{Umetsu2017}. In contrast, the single power-law, cored isothermal, and Burkert models are statistically disfavored by the averaged lensing profile having a pronounced radial curvature.

For each of the 20 clusters, \citet{Umetsu2016} performed a spherical NFW fit to the reconstructed projected mass density profile by accounting for all relevant sources of uncertainty, including measurement errors, cosmic noise due to the projection of large-scale structure uncorrelated with the cluster, statistical fluctuations of the projected cluster lensing signal due to halo triaxiality and correlated substructures.

In this analysis, we use the CLASH-RAR data set \citep{Tian2020} published in \citet{RARdata}, which contains $N_\mathrm{data}=84$ data points in $\log\gtot$--$\log\gbar$ space. \citet{Tian2020} extracted total and baryonic mass estimates where possible at $r=100$~kpc, $200$~kpc, $400$~kpc, and $600$~kpc. For each cluster, they also included a single constraint at $r\simlt 30$~kpc in the central BCG region. These data points are sufficiently well separated from each other, so as to avoid oversampling and reduce correlations between adjacent data points. 

The CLASH measurements of centripetal accelerations ($\gbar,\gtot$) are shown in  Figure~\ref{fig:CLASH}, along with our BAHAMAS predictions of the RAR for massive cluster halos with $E(z)M_{200}>5\times 10^{14}M_\odot$, derived from five different dark matter runs at $z=0.375$. The best-fit CLASH RAR obtained by \cite{Tian2020} has ${b_1}=0.51^{+0.04}_{-0.05}$, ${b_0}=-4.26^{+0.46}_{-0.47}$, and {$\Delta_\mathrm{int}$}$=0.064^{+0.013}_{-0.012}$ dex, which is in excellent agreement with the best-fit RAR for the BAHAMAS-CDM run characterized in the high-acceleration region of $\gbar>10^{-10.6}$~m~s$^{-2}$ (Table~\ref{tab:slope}).

\subsection{Statistical Comparison}
\label{subsec:comparison}

\begin{figure*}
 \centering
 \includegraphics[width=0.8\textwidth,clip]{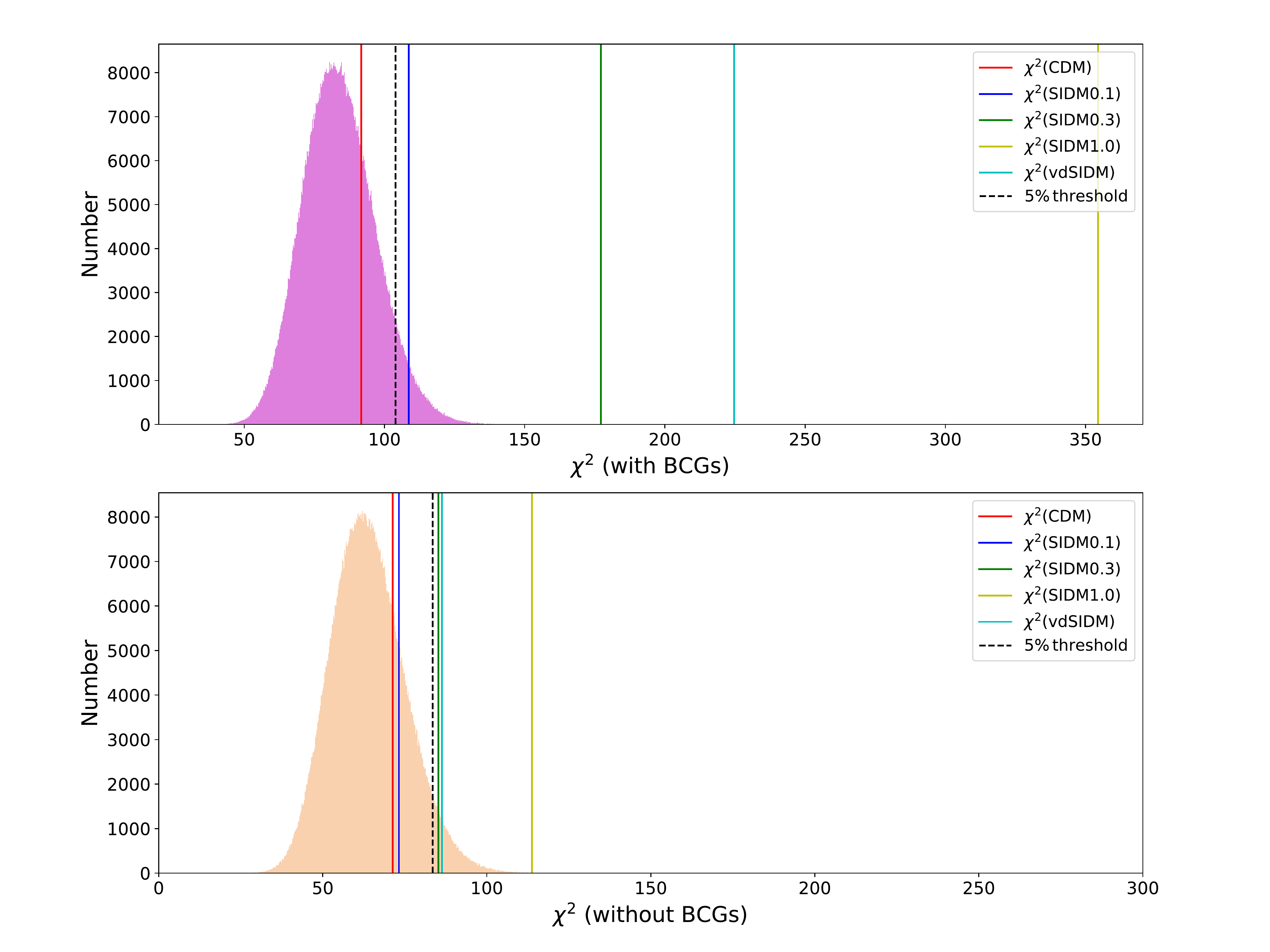}
 \caption{The $\chi^2$ distribution of the CLASH-RAR data set constructed from Monte-Carlo simulations. Solid vertical lines indicate the observed $\chi^2$ values for five different dark matter models. The black dashed vertical line corresponds to the significance threshold of $\alpha=0.05$. The upper and lower panels show the results with and without including the central $100$~kpc region, respectively.}
 \label{fig:chi2}
\end{figure*}

\begin{deluxetable*}{lcccccc}[tbp]
\centering
\tabletypesize{\footnotesize}
\tablecaption{Summary of the $\chi^2$ test using the $\gtot$--$\gbar$ relation (RAR)}
\label{tab:chi2}
\tablehead{
 \colhead{} & 
 \multicolumn{3}{c}{With BCGs} & 
 \multicolumn{3}{c}{Without BCGs} \\
 \cline{2-7} 
 \colhead{} & 
 \colhead{$\chi^2$ \tablenotemark{a}} &
 \colhead{PTE \tablenotemark{b}} &
 \colhead{Fraction \tablenotemark{c}} &
 \colhead{$\chi^2$ \tablenotemark{a}} &
 \colhead{PTE \tablenotemark{b}} &
 \colhead{Fraction \tablenotemark{c}} 
}
\startdata
 CDM      &  91.7 & 0.264  & 0.264 & 71.3 & 0.246 & 0.246\\
 SIDM0.1  & 108.6 & 0.037  & 0.036 & 73.2 & 0.200 & 0.200\\
 SIDM0.3  & 177.1 & $1.3\times10^{-8}$ & 0.0  & 85.3 & 0.039 &0.039\\
 SIDM1.0  & 354.5 & $6.3\times10^{-35}$ & 0.0 & 113.8 & $1.3\times10^{-5}$ & $1.4\times10^{-5}$\\
 vdSIDM   & 224.7 & $8.9\times10^{-15}$ & 0.0 & 86.3 & 0.033 & 0.033
\enddata
\tablenotetext{a}{Observed $\chi^2$ value between the CLASH data and each dark matter model.}
\tablenotetext{b}{Probability to exceed the observed $\chi^2$ value assuming the standard $\chi^2$ probability distribution function.}
\tablenotetext{c}{Fraction of Monte-Carlo realizations exceeding the observed value of $\chi^2$.}
\end{deluxetable*}

To make a quantitative comparison between the BAHAMAS simulations and the CLASH observations, we define the $\chi^2$ function as
\begin{equation}
 \chi^2 = \sum_{i=1}^{\Ndata}\frac{\left(\log_{10} g_{\mathrm{tot},i} - \log_{10} \widehat{g}_{\mathrm{tot}.i} \right)^2}{{\Delta_i^2} + {\Delta_\mathrm{int}^2}},
  \label{eq:chi2}
\end{equation}
where $i$ runs over all data points from the CLASH data set, $g_{\mathrm{tot},i}$ is the total acceleration at the $i$th data point, {$\Delta_i$} is the measurement uncertainty,  {$\Delta_\mathrm{int}$}$\approx 0.064$ is the intrinsic scatter for the CLASH sample determined by \citet{Tian2020}, and $\widehat{g}_{\mathrm{tot},i}$ is the theoretical prediction for the total acceleration at $g_\mathrm{bar}=g_{\mathrm{bar},i}$ from the BAHAMAS simulations. The resulting $\chi^2$ values evaluated for different dark matter runs are listed in Table~\ref{tab:chi2}.

To statistically characterize the level of agreement between data and simulations, we use frequentist measures of statistical significance. Specifically, we use the chance probability of exceeding the observed $\chi^2$ value to quantify the significance of the match for each dark matter run. For each case, we calculate the probability to exceed (PTE), or the right-tailed $p$-value, for a given value of $\chi^2$ assuming the standard $\chi^2$ probability distribution function. We adopt a significance threshold of $\alpha=0.05$ as the dividing line between satisfactory ($\mathrm{PTE}>0.05$) and unsatisfactory ($\mathrm{PTE}<0.05$) matches to the CLASH-RAR data set. Since no optimization (or model fitting) is performed in our $\chi^2$ evaluations, the number of degrees of freedom is $\Ndata=84$ in all cases. The resulting values of PTE for each dark matter run are listed in Table~\ref{tab:chi2}. We find that the CDM run ($\mathrm{PTE}=0.264$) gives a satisfactory match, whereas all the SIDM runs give unacceptable matches to the CLASH data. Among the SIDM runs, SIDM0.1 has a PTE of $0.036$ that is close to but slightly below the adopted threshold.

As a consistency check, we perform Monte-Carlo simulations to derive the $\chi^2$ distribution expected from the measurement errors $\{{\Delta_i}\}_{i=1}^{\Ndata}$ and the intrinsic scatter ${\Delta_\mathrm{int}}$ for the CLASH RAR. In each simulation, we construct a synthetic data set $\{g_{\mathrm{bar},i}^{(\mathrm{MC})}, g_{\mathrm{tot},i}^{(\mathrm{MC})}\}_{i=1}^{\Ndata}$  by creating a Monte-Carlo realization of random Gaussian noise $n_i \equiv \log_{10}g_{\mathrm{tot},i}^{(\mathrm{MC})} -\log_{10}\widehat{g}_{\mathrm{tot},i} = \pm\sqrt{{\Delta_i}^2 + {\Delta_\mathrm{int}^2}}$ at $\log_{10} g_{\mathrm{bar},i}^{(\mathrm{MC})}=\log_{10} g_{\mathrm{bar},i}^{(\mathrm{CLASH})}$ and then calculate the value of $\chi^2=\sum_i n_i^2/({\Delta_i^2+\Delta_\mathrm{int}^2})$. We repeat this procedure $10^6$ times to generate a large set of Monte-Carlo realizations and obtain the distribution of $\chi^2$ values. In Table~\ref{tab:chi2}, we list the fraction of Monte-Carlo realizations exceeding the observed value of $\chi^2$ for each dark matter run. In all cases, the Monte-Carlo fraction of exceeding the $\chi^2$ value is precisely consistent with the PTE calculated with the standard $\chi^2$ distribution function. In the upper panel of Figure~\ref{fig:chi2}, we compare the $\chi^2$ distribution of the CLASH data set constructed from our Monte-Carlo simulations with the observed $\chi^2$ values for the CDM and four SIDM models. This figure gives a visual summary of the $\chi^2$ test (Table~\ref{tab:chi2}).

\begin{deluxetable}{lcccc}[tbp]
\centering
\tabletypesize{\footnotesize}
\tablecaption{Likelihood-ratio test of the SIDM models}
\label{tab:sigma}
\tablehead{
 \colhead{} & 
 \multicolumn{2}{c}{With BCGs} & 
 \multicolumn{2}{c}{Without BCGs}\\ 
 \cline{2-5} 
 \colhead{} & 
 \colhead{$\Delta\chi^2$} &
 \colhead{Significance level} &
 \colhead{$\Delta\chi^2$} &
 \colhead{Significance level} 
}
\startdata
SIDM0.1  & 16.9  & 4.1$\sigma$& 1.9    &1.4$\sigma$\\
SIDM0.3  & 85.4  & 9.2$\sigma$ & 14.0  &3.8$\sigma$\\
SIDM1.0  & 262.8 & 16.2$\sigma$ & 42.5 & 6.8$\sigma$\\
vdSIDM   & 133.0 & 11.2$\sigma$ & 15.0 & 3.4$\sigma$
\enddata
\tablecomments{The observed value of $\Delta\chi^2=\chi^2-\chi^2_\mathrm{CDM}$ relative to the CDM model is listed for each SIDM model explored in this work. The number of degrees of freedom for each comparison is 1 for all cases except for vdSIDM with 2 degrees of freedom.}
\end{deluxetable}

We perform a likelihood-ratio test of SIDM models to quantify whether the inclusion of collisional features of dark matter is statistically warranted by the data. Velocity-independent SIDM models have one additional degree of freedom relative to the CDM model. For vdSIDM, there are two additional degrees of freedom relative to the CDM model. These models are reduced to the CDM model in the limit of the vanishing cross-section. Table~\ref{tab:sigma} lists for each SIDM model the differences in the $\chi^2$ value $\Delta\chi^2=\chi^2-\chi^2_{\mathrm{CDM}}$ relative to the CDM model and the corresponding significance level. Compared with the fiducial model of CDM, SIDM0.1 is disfavored at a significance level of $4.1\sigma$. 

These results are based on the CLASH RAR at BCG--cluster scales, which includes, for each cluster, a single constraint in the central BCG region at $r\simlt 30$~kpc. However, since the typical resolution of CLASH mass reconstructions is $\Delta r\approx 50$~kpc (Section~\ref{subsec:clash}), the mass distribution is not resolved in the BCG region and thus the CLASH constraints on $\gtot$ at the BCG scale are model dependent to some extent. Moreover, the distribution of baryonic and dark matter in cluster cores is sensitive to baryonic physics \citep[e.g.,][]{Cui2018}. 

We therefore repeat the tests described above using core-excised RAR data. Excluding the central $r<100$~kpc region from the CLASH data set, we have 64 data points at $r \in [100,600]$~kpc. The results of the $\chi^2$ test with the core-excised data set are summarized in Table~\ref{tab:chi2}. Of the five BAHAMAS dark matter runs, CDM and SIDM0.1 provide satisfactory matches to the core-excised CLASH RAR, at a significance level of $\alpha=0.05$. Excluding the BCG regions results in a weaker but still competitive constraint on the SIDM cross-section (Table~\ref{tab:sigma}). With a likelihood ratio test, we find that the core-excised CLASH RAR data disfavor the SIDM0.3 model at the $3.8\sigma$ level with respect to the CDM model.

We note that in the above analysis, the measurement uncertainty of $\log_{10}\gbar$ (which is typically $\sim 7\%$ of that of $\log_{10}\gtot$ and thus negligible in the analysis) is not taken into account. In Appendix~\ref{app_2}, we perform an alternative analysis in terms of the $\gtot/\gbar$--$\gbar$ relation, which is referred to as the mass discrepancy--acceleration relation \citep[MDAR; see][]{Famaey2012}. In this MDAR analysis, we can explicitly account for the uncertainty in the $\gbar$ measurements. We find that the inclusion of the measurement uncertainty in $\gbar$ has only a minor impact on the statistical inference and does not change our conclusions regarding the acceptance of dark matter models.

\section{Discussion}
\label{sec:discussion}

In this section, we first discuss current limitations and possible improvements of SIDM constraints from measurements of the cluster RAR. Then, we compare our results to previous astrophysical constraints on SIDM.

\subsection{Possible Systematics and Improvements}

In this work, we analyzed the CLASH RAR data set of \cite{Tian2020}, which consists of $\Ndata=84$ data points  ($\gbar,\gtot$) inferred from the multiwavelength CLASH observations of 20 high-mass galaxy clusters (Section~\ref{subsec:clash}). In \citet{Tian2020}, the measurements of centripetal accelerations were sparsely sampled over a sufficiently wide range of clustercentric distances, so as to reduce the covariance between adjacent data for each cluster. Thus, our comparison of the BAHAMAS simulations and CLASH measurements (Equation~(\ref{eq:chi2})) involves a two step procedure, which ignores the covariance and does not fully exploit all the information contained in the data.  As a result, our analysis is likely to overestimate the significance of our constraints. In principle, these limitations can be overcome by using a forward-modeling method. In particular, likelihood-free approaches based on forward simulations allow us to bypass the need for a direct evaluation of the likelihood function assuming Gaussian statistics, which avoids the complex derivation of the covariance matrix in an inherently complex problem \citep[e.g.,][]{Tam2022}.

Another potential source of systematic uncertainty is the smoothing of the inner density profile due to cluster miscentering \citep[e.g.,][]{Johnston2007}.  Because of the CLASH selection, our cluster sample exhibits, on average, a small positional offset between the BCG and X-ray peak, characterized by an rms offset of $\sim 40$~kpc \citep{Umetsu2014,Umetsu2016}. This level of offset is comparable to the typical effective radius $R_\mathrm{e}$ of CLASH BCGs ($\langle R_\mathrm{e}\rangle\sim 30$~kpc; Section~\ref{sec:observed}) but sufficiently small compared to the range of cluster radii of interest (say, $r\simgt 100$~kpc). Moreover, since the RAR method uses the same aperture to compare total and baryonic accelerations, the miscentering effect is not expected to significantly affect the SIDM constraint, although it could potentially contribute to the scatter in the RAR inferred from cluster observations.

The total and baryonic mass profiles of galaxy clusters, $\Mtot(<r)$ and $\Mbar(<r)$, inferred from observational data are model dependent to some extent. \citet{Tian2020} found that the observed surface brightness distribution of CLASH BCGs is well described by a de Vaucouleurs’ profile, which is a special case of the Sersic model, with a Sersic index of $n=4$. Accordingly, they modeled the 3D stellar mass distribution in the CLASH BCGs with a Hernquist profile, which closely resembles the de Vaucouleurs’ law in projection \citep[][see their Figure 4]{1990ApJ...356..359H}, especially at $\simgt 0.5 R/R_{\mathrm{e}}$ where \cite{Tian2020} extracted stellar mass estimates of the CLASH BCGs.

On the other hand, \citet{Tian2020} modeled the total mass distribution of CLASH clusters with an NFW profile, which best describes the stacked lensing profile of the CLASH sample \citep[][see Section~\ref{subsec:clash}]{Umetsu2016}. However, the choice of the cuspy NFW model implicitly assumes collisionless CDM, which could lead to a biased inference of the cluster RAR in an SIDM cosmology. In particular, this NFW assumption in our analysis is likely to underestimate the goodness of fit for the SIDM models. In future studies, it will thus be important to consider more flexible and self-consistent mass models, such as the Einasto \citep{Einasto} and Diemer--Kravtsov \citep{DK14} density profiles, with an additional parameter to capture the radial curvature of the central density profile that depends on the SIDM cross-section \citep{Eckert2022}. 

\subsection{Comparison with Other Work}

Self-interactions between dark matter particles are expected to make halos more spherical compared to triaxial halos of collisionless CDM, especially in the central region where the scattering rate is largest. Self-interactions in the optically thin regime also reduce the central dark matter densities, transforming a cusp into a core. Moreover, offsets between the galactic and dark-matter centroids in merging clusters can be used to constrain the SIDM cross-section. Previous studies placed upper limits on the self-interaction cross-section of dark matter particles using such observed density features in galaxy clusters \citep[for a review, see][]{Tulin+Yu2018}. 

\cite{2013MNRAS.430..105P} compared the halo ellipticities inferred from lensing and X-ray observations with cosmological simulations with SIDM cross-sections of $\sigma/m=0.03$, $0.1$, and $1$~cm$^2$~g$^{-1}$. They found that the strong-lensing measurement of the cluster ellipticity for MS~2137−23 \citep{Miralda-Escude2002} is compatible with an SIDM cross-section of $\sigma/m= 1$~cm$^2$~g$^{-1}$, whereas the X-ray shape measurement of the isolated elliptical galaxy NGC~720 \citep{Buote2002} is consistent with $\sigma/m= 0.1$~cm$^2$~$g^{-1}$. 

Using the galaxy--dark-matter offset measured in the moving subcluster of the Bullet Cluster, \citet[][]{2008ApJ...679.1173R} placed an upper limit of $\sigma/m < 1.25$~cm$^{2}$~g$^{-1}$ at the $68\%$~CL. \cite{2015Sci...347.1462H} performed an ensemble analysis of offset measurements for 72 substructures in 30 systems, including both major and minor mergers, and set an upper limit of $\sigma/m<0.47$~cm$^2$~g$^{-1}$ at the $95\%$~CL. \cite{2018ApJ...869..104W} revisited the analysis of \cite{2015Sci...347.1462H} using more comprehensive data and carefully reinterpreted their refined offset measurements, finding that the SIDM constraint of \cite{2015Sci...347.1462H} is relaxed to $\sigma/m\lesssim 2$~cm$^2$~g$^{-1}$.

Analyzing X-ray and SZ effect observations, \citet{Eckert2022} constrained the structural parameters of the mass density profiles for a sample of 12 massive X-COP clusters assuming that the intracluster gas is in hydrostatic equilibrium. They used the BAHAMAS-SIDM simulations to construct an empirical scaling relation between the Einasto shape parameter and the velocity-independent SIDM cross-section. With this relation and the assumption of hydrostatic equilibrium, they obtained an upper limit of $\sigma/m<0.19$~cm$^{2}$~g$^{-1}$ at the $95\%$~CL.
 
In this study, we have obtained competitive constraints on the SIDM cross-section using cluster RAR measurements. By comparing the CLASH RAR data set with the mean RARs derived from the BAHAMAS simulations, we are able to reject an SIDM model with $\sigma/m=0.1$~cm$^{2}$~g$^{-1}$ at the $4.1\sigma$~CL with respect to the CDM model. Excluding the central $r<100$~kpc region, we find that an SIDM model with $\sigma/m=0.3$~cm$^{2}$~g$^{-1}$ is disfavored at the $3.8\sigma$ level with respect to CDM.

\section{Summary and Conclusions} 
\label{sec:summary}

The RAR is a tight relation between the total and baryonic centripetal accelerations, $\gtot$ and $\gbar$, inferred for galaxies and galaxy clusters \citep{2016PhRvL.117t1101M,Tian2020}. This tight empirical correlation offers a new possibility of testing the collisionless nature of dark matter at galaxy--cluster scales.

As a first step toward this goal, we studied in this paper the RAR in simulated halos for both CDM and SIDM models, using the BAHAMAS suite of cosmological hydrodynamical simulations \citep[][]{2017MNRAS.465.2936M,2018MNRAS.476.2999M,2019MNRAS.488.3646R}. We analyzed simulations at $z=0$ and $z=0.375$ run with four different SIDM models (SIDM0.1, SIDM0.3, SIDM1.0, and vdSIDM; see Section~\ref{sec:data}), as well as collisionless CDM. 

For each dark matter model, we have determined the mean $\gtot$ as a function of $\gbar$, or the halo RAR, in halos of different mass bins (Section~\ref{sec:result}; see Figure~\ref{fig:all_RAR}). We find that the slope of the halo RAR at high acceleration ($\gbar\simgt 10^{-10.6}$~m~s$^{-2}$) decreases with increasing SIDM cross-section. This flattening feature at high $\gbar$ is more significant in more massive halos at lower redshift (Figures~\ref{fig:all_RAR_mass_split} and \ref{fig:compare_z}), consistent with the fact that the scattering rate is proportional to the local dark matter density and velocity dispersion (Equation~(\ref{eq:scatter})). This suggests that the high-$\gbar$ cluster-scale RAR for low-redshift samples can be used to probe the nature of dark matter. 

Focusing on massive cluster halos at $z=0.375$, we have also characterized the slope (${b_{1}}$), intercept (${b_{0}}$), and intrinsic scatter (${\Delta_\mathrm{int}}$) of the mean RAR at high $\gbar$ for different dark matter models (see Table~\ref{tab:slope}). In all cases, we find a remarkably tight distribution in $\log\gtot$--$\log\gbar$ space, with a slight increase in ${\Delta_\mathrm{int}}$ with increasing SIDM cross-section. We find that only the CDM case can be well described by Equation~(\ref{eq:McGaugh}) proposed by \citet{2016PhRvL.117t1101M}, with an acceleration scale of $g_{\dagger} = (1.42\pm0.06)\times 10^{-9}$~m~s$^{-2}$. This is much higher than the characteristic acceleration scale of $g_{\dagger}\approx 1.2\times 10^{-10}$~m~s$^{-2}$ observed at galaxy scales \citep{2016PhRvL.117t1101M}.

We have compared the halo RARs from the BAHAMAS-CDM and -SIDM runs to the cluster RAR inferred from CLASH observations (Section~\ref{sec:observed}; see Tables~\ref{tab:chi2} and \ref{tab:sigma}). Our comparison shows that the RAR in the CDM model provides an excellent match to the CLASH RAR \citep{Tian2020}. This comparison includes the high-$\gbar$ regime probed by the BCGs. By contrast, models with a larger SIDM cross-section (hence with a greater flattening in $\gtot$ at high $\gbar$) yield increasingly poorer matches to the CLASH RAR. Excluding the BCG regions, we obtain a weaker but still competitive constraint on the SIDM cross-section. Using the RAR data outside the central $r<100$~kpc region, we find that an SIDM model with $\sigma/m=0.3$~cm$^{2}$~g$^{-1}$ is disfavored at the $3.8\sigma$ level with respect to the CDM model. However, it should be noted that the choice of the NFW model used for fitting the CLASH lensing data \citep{Tian2020} implicitly assumes collisionless CDM, which could lead to a biased inference of the cluster RAR in an SIDM cosmology. As a result, this NFW assumption is likely to underestimate the goodness of fit for the SIDM models. In future work, it will be important to use more flexible mass models with an additional parameter to describe the central density slope that depends on the SIDM cross-section \citep{Eckert2022}.

In this study, we have demonstrated the power and potential of the RAR for testing the collisionless nature of dark matter. Thus far, the cluster RAR has been determined using gravitational lensing only for the CLASH sample. To place stringent and robust constraints on the SIDM cross-section, it is necessary to increase the sample of clusters for which lensing and X-ray observations are available over a broad radial range down to $r\sim 100$~kpc (Section~\ref{subsec:massdep}). For nearby clusters at $z<0.1$, such mass measurements can be obtained from wide-field weak-lensing observations \citep[e.g.,][]{Okabe2014}. For clusters at higher redshifts, combined strong and weak lensing is required to distinguish SIDM and CDM using the RAR. The ongoing CHEX-MATE project will provide such ideal multiwavelength data sets of high quality, for a minimally biased, signal-to-noise-limited sample of 118 Planck galaxy clusters at $0.05<z<0.6$ detected through the SZ effect \citep{CHEXMATE2021}. Extending this work to the CHEX-MATE sample will thus be a substantial step toward understanding the collisionless nature of dark matter.



\acknowledgments

We thank the anonymous referee for valuable comments that improved the clarity of the paper.
We acknowledge fruitful discussions with Yong Tian, Teppei Okumura, and Stefano Ettori. This work is supported by the Ministry of Science and Technology of Taiwan (grant MOST 109-2112-M-001-018-MY3) and by the Academia Sinica Investigator award (grants AS-IA-107-M01 and AS-IA-112-M04). Parts of this research were carried out at the Jet Propulsion Laboratory, California Institute of Technology, under a contract with the National Aeronautics and Space Administration (80NM0018D0004).

\software{Astropy \citep{2018AJ....156..123A}, matplotlib \citep{Matplotlib}, NumPy \citep{numpy}, Python \citep{python3}, Scipy \citep{scipy}}

\clearpage

\begin{appendix}

\section{Correlation between the Total and Baryonic Matter Distributions}
\label{app_1}

The key features used to distinguish different dark matter models mainly come from the inner region of cluster halos ($r\simlt 200$~kpc), as shown in Figure~\ref{fig:all_RAR_radial_cut}. Here, we further investigate whether the correlation between the acceleration ratio $\gtot/\gbar$ (or the total-to-baryonic mass ratio, $\Mtot/\Mbar$) and the clustercentric distance $r$  can provide sufficient information to distinguish between different dark matter models.

\begin{deluxetable}{lcc}
\centering
\tablecaption{ Logarithmic scatter around the mean $\gtot/\gbar$--$r$ relation evaluated in the inner region of cluster-scale halos}
\label{tab:scatter_ggg}
\tablehead{
 \colhead{Dark matter model} & 
 \colhead{$\sigma(<400~\mathrm{kpc})$} &
 \colhead{$\sigma(<200~\mathrm{kpc})$}
 \\
 \colhead{} &
 \colhead{(dex)} &
 \colhead{(dex)}
 }
\startdata
CDM &    0.104 & 0.111\\
SIDM0.1 &  0.104 & 0.113\\
SIDM0.3 &  0.141 & 0.156\\
SIDM1.0  &0.138 & 0.154\\
vdSIDM &   0.125 & 0.137
\enddata
\tablecomments{ For each model, the $\gtot/\gbar$--$r$ relation is derived for a subsample of cluster halos at $z=0.375$ with masses $E(z)M_{200}>5\times 10^{14}M_\odot$.  An inner cut-off radius of $\rcut=15$~kcp is used.}
\end{deluxetable}

\begin{figure}[tbp]
 \centering
 \includegraphics[width=0.45\textwidth,clip]{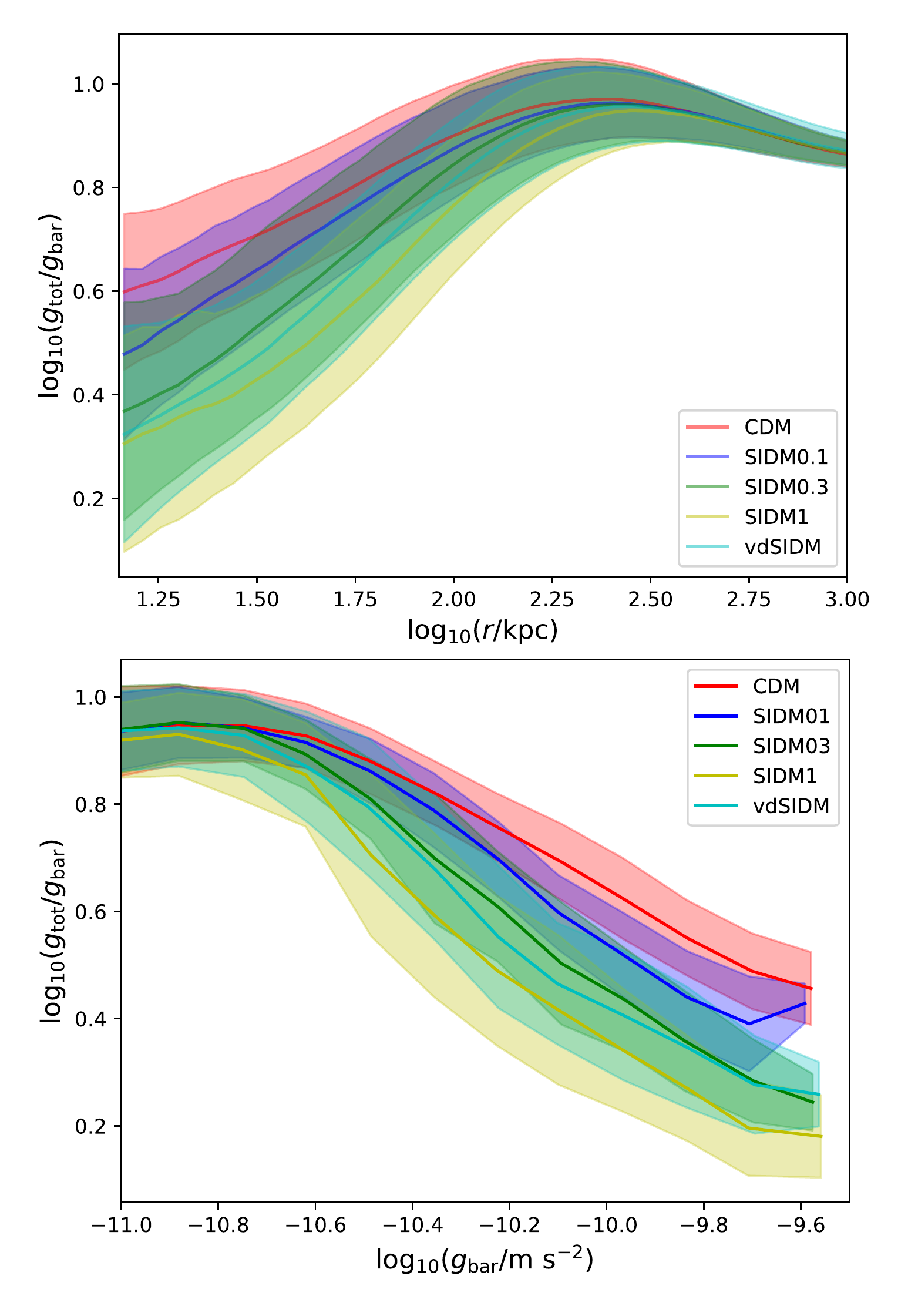}
 \caption{ Ratio of the total acceleration $\gtot$ to the baryonic acceleration $\gbar$ as a function of cluster radius $r$ (upper panel) and as a function of baryonic acceleration $\gbar$ (lower panel) for massive cluster-scale halos with $E(z)M_{200}>5\times 10^{14}M_\odot$ at $z=0.375$. The results are shown separately for five different dark matter simulation runs. For each dark matter model, the solid line represents the mean relation and the shaded area shows the standard deviation around the mean. }
 \label{fig:gtotgbar_r}
\end{figure}

{The top panel of Figure~\ref{fig:gtotgbar_r} shows the mean and standard deviation profiles of $\gtot(r)/\gbar(r)$ as a function of cluster radius $r$, derived for each of the dark matter runs of the BAHAMAS simulation at $z=0.375$ using the high-mass subsample with $E(z)M_{200}>5\times 10^{14}M_\odot$. The values of the logarithmic scatter of $\gtot(r)/\gbar(r)$ evaluated in the inner region of cluster halos ($r< 400$~kpc or $r<200$~kpc) are listed in Table~\ref{tab:scatter_ggg}.  The radial range of $r<400$~kpc corresponds approximately to the high-acceleration region with $\gbar > 10^{-10.6}$~m~s$^{-2}$ used to characterize the intrinsic scatter around the mean RAR, or the $\gbar$--$\gtot$ relation (Table~\ref{tab:slope}). 

Overall, the magnitude of the scatter in the $\gtot/\gbar$--$r$ relation is about a factor of $\sim 1.6$ larger than that of the RAR (see Table~\ref{tab:slope}). As a result, the radial $\gtot(r)/\gbar(r)$ profiles of cluster halos derived from the different dark matter runs significantly overlap with each other, even in their innermost region (Figure~\ref{fig:gtotgbar_r}). Therefore, we conclude that compared to the RAR or MDAR, the $\gtot/\gbar$--$r$ relation (i.e., the $\Mtot(r)/\Mbar(r)$ profile) has a much weaker sensitivity to the SIDM cross-section.}

\begin{figure*}[tbp]
 \centering
 \includegraphics[width=0.7\textwidth,clip]{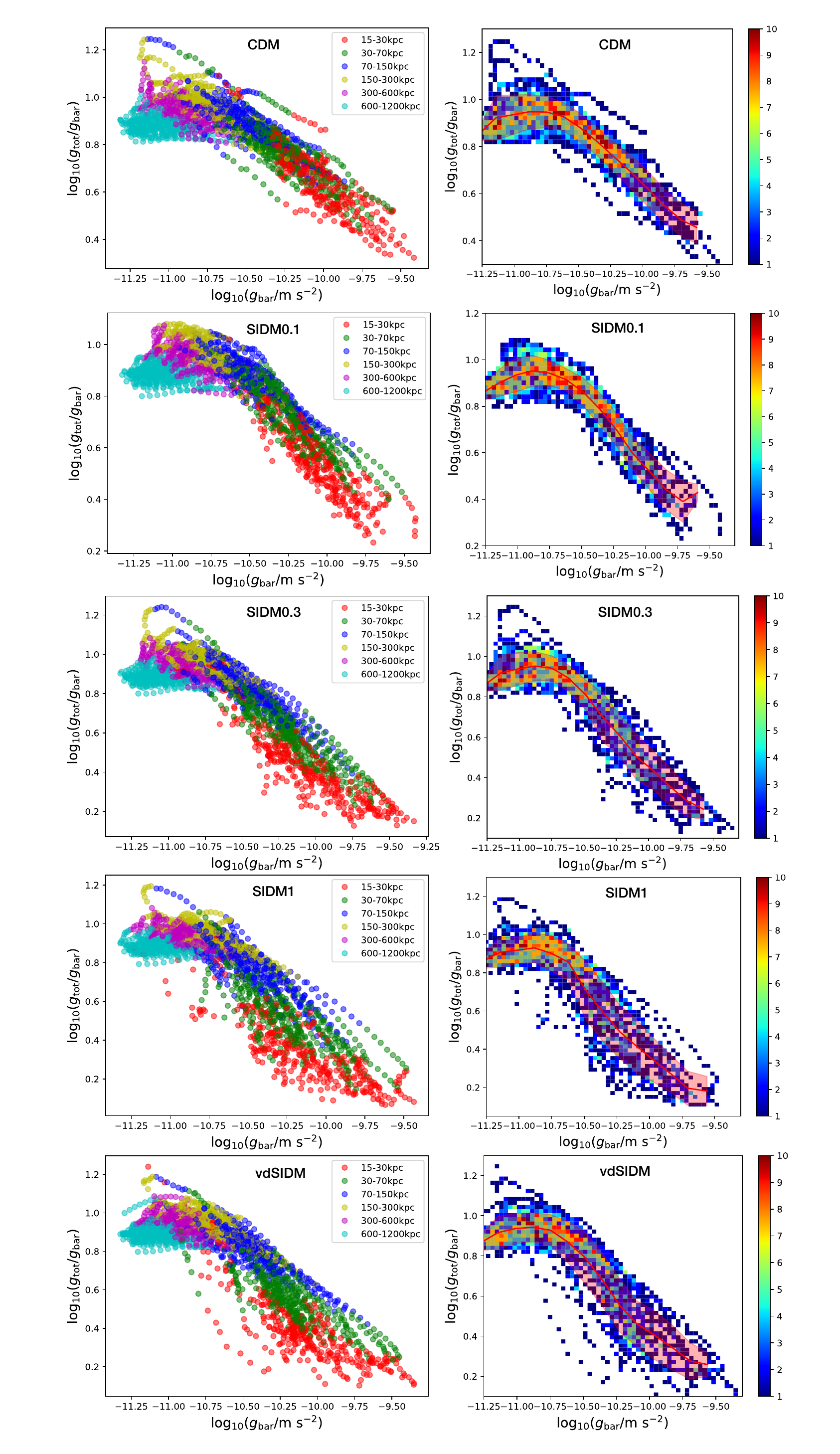}
 \caption{The $\gtot/\gbar$--$\gbar$ diagram derived for a subsample of massive cluster-scale halos with $E(z)M_{200} > 5\times 10^{14}M_\odot$ at $z=0.375$. The results are shown separately for five different dark matter runs of the BAHAMAS simulation.  In the left panels, the data are divided into six different radial bins ranging from $r=15$~kpc to $1200$~kpc. All points in each panel are color-coded according to their radial bin. In the right panels, the corresponding histogram distribution is shown for each dark matter run. In each panel, the red solid line represents the mean relation and the red shaded area show the standard deviation around the mean relation.}
 \label{fig:gtotgbar_gbar}
\end{figure*}

{To understand the tightness of RAR/MDAR, we split the simulation data in $\log(\gtot/\gbar)$--$\log\gbar$ space into six different radial bins ranging from $15$~kpc to $1200$~kpc and show them separately in the left panels of Figure~\ref{fig:gtotgbar_gbar}. In each radial bin, the MDAR sequence spans a wide range in $\gbar$, indicating that even dark matter particles at large clustercentric distances can contribute to the high acceleration region. The distribution of $\gtot/\gbar$ in a given radial bin is obtained by projecting each set of color-coded data onto the $y$-axis in Figure~\ref{fig:gtotgbar_gbar}, which results in a large scatter around the mean value $\langle \gtot/\gbar\rangle$. Therefore, the tight correlation in the RAR/MDAR appears to be a unique signature of the empirical coupling between the total and baryonic components, which increases the ease of distinguishing between CDM and SIDM models.

The physical origin and nature of this tight empirical coupling between the total and baryonic accelerations are still under debate. In the future, it will be useful to search for different combinations of the total and baryonic mass properties of galaxy clusters to identify other low-scatter probes of the SIDM cross-section.} 

{

\section{Accounting for the measurement uncertainty of the baryonic acceleration}
\label{app_2}

\begin{figure}[bp]
 \centering
 \includegraphics[width=0.5\textwidth,clip]{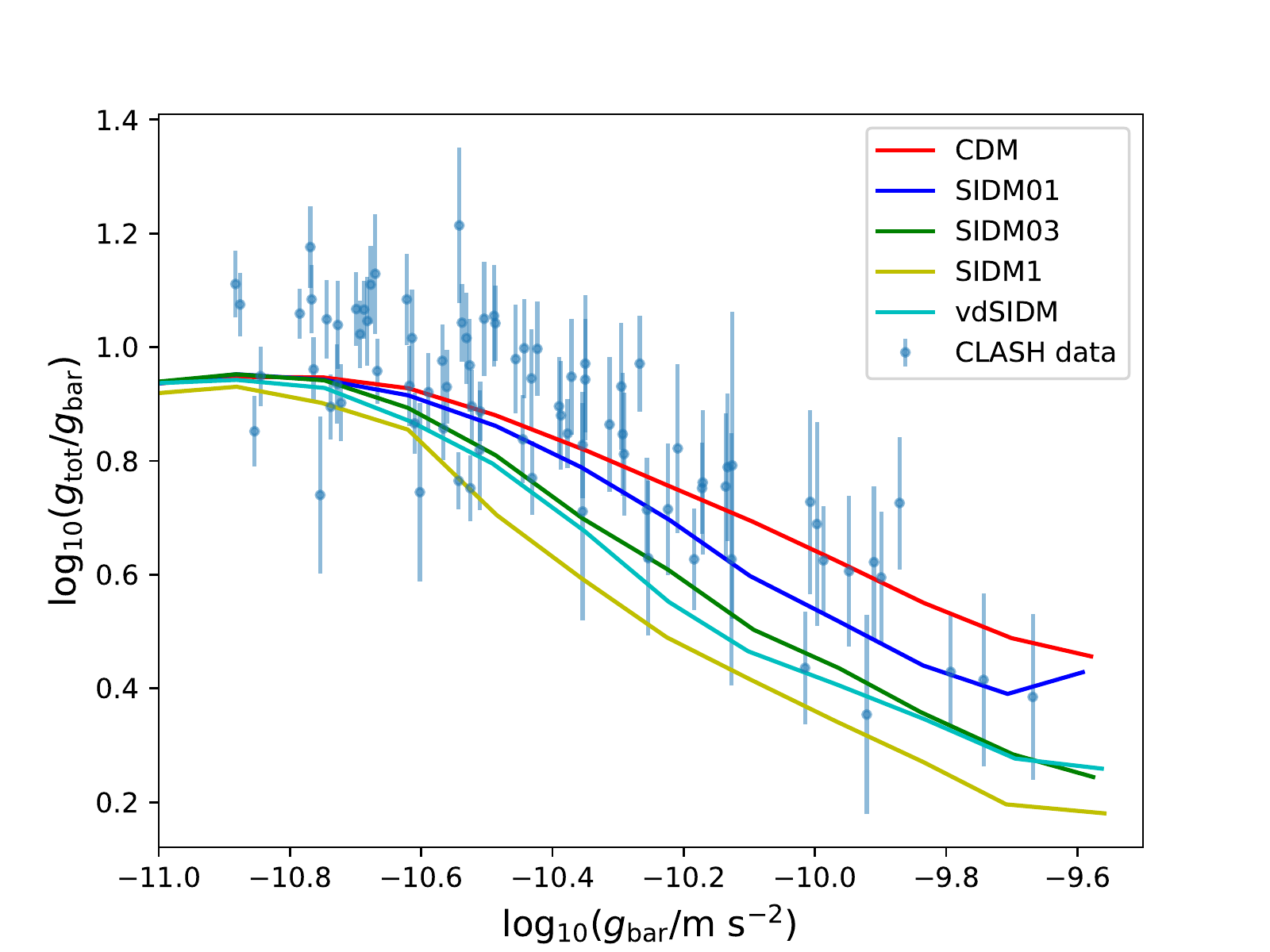}
 \caption{Comparison of the CLASH data with our BAHAMAS predictions of the cluster-scale $\gtot/\gbar$--$\gbar$ relation at $z=0.375$, derived from five different dark matter runs (thick color lines). The CLASH measurements are shown with blue points with error bars.}
 \label{fig:CLASH_a1}
\end{figure}

In this appendix, we explicitly account for the measurement uncertainty of $\gbar$ in our analysis of the CLASH RAR data set. To this end, we utilize the MDAR, namely $\gtot/\gbar$ as a function of $\gbar$, and follow the analysis procedure described in Section~\ref{subsec:comparison}. The $\chi^2$ function in this analysis is defined by
\begin{equation}
\begin{gathered}
 \chi^2 = \sum_{i=1}^{\Ndata}\frac{\left[ \log_{10}(g_{\mathrm{tot},i}/g_{\mathrm{bar},i}) - \log_{10} (\widehat{g}_{\mathrm{tot},i}/\widehat{g}_{\mathrm{bar},i}) \right]^2}{\Delta_i^2 + \Delta_\mathrm{int}^2},
  \label{eq:chi2_R}
\end{gathered}
\end{equation}
where $i$ runs over all data points from the CLASH data set, $\Delta_{i}^{2}=\Delta^{2}(\log_{10}g_{\mathrm{tot},i})+\Delta^{2}(\log_{10}g_{\mathrm{bar},i})$ represents the total uncertainty including the measurement errors in both $\log_{10}\gbar$ and $\log_{10} \gtot$, and $\Delta_\mathrm{int}$ is the lognormal intrinsic scatter of $\gtot$ at fixed $\gbar$.

Figure~\ref{fig:CLASH_a1} shows the distribution of the CLASH-MDAR data and the BAHAMAS predictions of the mean $\gtot/\gbar$--$\gbar$ relation derived from five different dark matter runs. The resulting $\chi^2$ values and corresponding PTE values evaluated for the respective dark matter runs are listed in Table~\ref{tab:chi2_R}.

\begin{deluxetable}{lcccccc}[tbp]
\centering
\tabletypesize{\footnotesize}
\tablecaption{Summary of the $\chi^2$ test using the $\gtot/\gbar$--$\gbar$ relation (MDAR)}
\label{tab:chi2_R}
\tablehead{
 \colhead{} & 
 \multicolumn{2}{c}{With BCGs} & 
 \multicolumn{2}{c}{Without BCGs} \\
 \cline{2-5} 
 \colhead{} & 
 \colhead{$\chi^2$\tablenotemark{a}} &
 \colhead{PTE\tablenotemark{b} } &
 \colhead{$\chi^2$\tablenotemark{a} } &
 \colhead{PTE\tablenotemark{b} } &
}
\startdata
 CDM      & 89.32 & 0.314 & 70.18 & 0.278  \\
 SIDM0.1  & 106.8 & 0.047 & 73.46 & 0.221\\
 SIDM0.3  & 177.4 & $1.2\times10^{-8}$ & 85.89 & 0.035\\
 SIDM1.0  & 348.5 & 0.0 & 112.9 & $1.6\times10^{-4}$\\
 vdSIDM   & 221.0 & $2.9\times10^{-14}$ & 83.65 & 0.031
\enddata
\tablenotetext{a}{Observed $\chi^2$ value between the CLASH data and each dark matter model.}
\tablenotetext{b}{Probability to exceed the observed $\chi^2$ value assuming the standard $\chi^2$ probability distribution function.}
\end{deluxetable}

When the central BCG constraint is included, only the CDM model gives a satisfactory match to the CLASH data set, which is consistent with the result of the RAR-based analysis given in Section~\ref{subsec:comparison}. However, we note that the SIDM0.1 model now has a PTE value of 0.047, which is very close to but slightly below the significance threshold of $\alpha=0.05$. When excluding the central BCG region of $r<100$~kpc, we find that both CDM and SIDM0.1 provide satisfactory matches to the CLASH data. All these results are consistent with the findings based on the RAR analysis presented in Section~\ref{subsec:comparison}, suggesting that the inclusion of the measurement uncertainty of $\gbar$ has only a minor impact on the statistical inference and it does not change our conclusions of the accepted dark matter models.}

\end{appendix}

\bibliography{sample63}

\end{document}